\documentclass[amsmath,longbib,raggedbottom]{aastex702}
\usepackage{mathtools,bm,mathrsfs,booktabs,placeins,enumitem}
\allowdisplaybreaks[1]
\shorttitle{Satellite Survival by Gas Redistribution}
\shortauthors{Mosqueira}
\definecolor{referenceblue}{RGB}{0,51,102}
\hypersetup{
  colorlinks=true,
  citecolor=referenceblue,
  urlcolor=referenceblue,
  linkcolor=black,
  pdfborder={0 0 0},
  pdftitle={Satellite Survival through Gas Redistribution in a Low-viscosity Disk},
  pdfauthor={I. Mosqueira}
}
\newcommand{\GEnd}{1,000}
\newcommand{\GFinalA}{16.1}
\newcommand{\GMinA}{15.1}
\newcommand{\GMaxE}{3.43\times10^{-5}}
\newcommand{\GMinSigma}{264}
\newcommand{\GFinalLocalSigma}{4.90\times10^{3}}
\newcommand{\GInitialRate}{-7.21\times10^{-2}}
\newcommand{\GLastRate}{-5.89\times10^{-5}}
\newcommand{\GLastReduction}{99.9}
\newcommand{\GExcitationRatio}{0.152}
\newcommand{\GBalanceRatio}{-3.26\times10^{-3}}
\newcommand{\GNetRatio}{-4.97\times10^{-4}}
\newcommand{\PairEnd}{547}
\newcommand{\PairGapRatio}{0.067}
\newcommand{\RafG}{13.2}
\newcommand{\RafC}{8.48}

\newcommand{\BuoyFinalA}{17.9}

\newcommand{\BuoyLocalSigma}{169}
\newcommand{\BuoyLocalPercent}{0.75}
\newcommand{\BuoyOuterPercent}{0.70}
\newcommand{\BuoyRMS}{0.0108}
\newcommand{\ControlRMS}{0.21}
\newcommand{\BuoyPtp}{0.0348}
\newcommand{\ControlPtp}{0.721}
\newcommand{\BuoyRMSFactor}{19.4}
\newcommand{\BuoyRateReduction}{99.5}

\newcommand{\BuoySnapshots}{498}
\newcommand{\BuoyOrbitStates}{10004}
\newcommand{\BuoyLateRate}{-2.52\times10^{-4}}
\newcommand{\BuoyMaxE}{2.39\times10^{-5}}
\newcommand{\BuoyMassResidual}{1.50\times10^{-11}}
\newcommand{\BuoyJResidual}{1.08\times10^{-11}}
\newcommand{\BuoyFullJPercent}{-0.075}

\newcommand{\BuoyLateReturnPercent}{77.5}
\newcommand{\BuoyAnnulusPercent}{3.43}
\newcommand{\BuoyAnnulusRefMass}{4.00\times10^{26}}
\newcommand{\LowFirstNet}{-4.86\times10^{31}}
\newcommand{\HighFirstNet}{5.20\times10^{31}}
\newcommand{\LowLateNet}{-4.33\times10^{31}}
\newcommand{\GradientLateRate}{3.50\times10^{-4}}
\newcommand{\CalibratedLateRate}{4.11\times10^{-4}}

\begin{document}
\title{Satellite Survival through Gas Redistribution in a Low-viscosity Disk}
\author[gname=Ignacio,sname=Mosqueira]{I. Mosqueira}
\affiliation{San Jos\'e State University, One Washington Square, San Jos\'e, CA
95192, USA}
\email{Ignacio.Mosqueira@sjsu.edu}

\begin{abstract}

We investigate satellite survival through gas redistribution in a dense
circumplanetary disk with a low dimensionless viscosity parameter
$\alpha \sim 10^{-6}$. The model
combines modal Lindblad excitation, launch-dependent shock deposition, 3D
effects, and an instantaneous Rayleigh adjustment that conserves angular momentum
and suppresses sharp density gradients. An isolated Ganymede-mass satellite depletes the
disk outside its orbit and stalls under a calibrated
three-dimensional Lindblad torque. This behavior is consistent with the
non-feedback branch of Rafikov's stalling criterion. The inward migration stalls
near 15 Jupiter radii ($R_J$) when the gas depletion exterior to the satellite's
orbit reduces the outer torque by the amount required 
to balance the torque of the undepleted
inner disk, even when we adopt a transport prescription that begins smoothing
density gradients halfway to Rayleigh marginality. Likewise, two Callisto
masses form an extended depleted region and stall in nearly steady
orbits before a late close encounter; however, a self-consistent disk 
response to satellite eccentricity remains to be modeled.
A simulation of a Ganymede-mass satellite adds a specified
source of unsaturated local angular momentum deposition by buoyancy
torques, reducing the late radial oscillations indirectly caused by non-local shock
deposition. Lastly, we redistribute gas while conserving angular momentum
to proactively inhibit Rossby-wave unstable pressure bumps at the gap edges. We find 
that this approach preserves gap clearing and strong migration suppression
for Ganymede and the two Callisto masses. 
In the pressure-adjusted run, the two Callisto masses are captured in a 7:5 resonance 
and retain eccentricities below 0.06.
\end{abstract}

\section{Satellite survival through gas disk redistribution}
\label{sec:introduction}

A large embedded satellite redistributes the gas that drives its migration. Weak
transport allows this redistribution to limit orbital loss in dense satellite
disks \citep{MosqueiraEstrada2003b}. Observations of PDS~70~c and SR~12~c support
quiescent, solids-enhanced satellite-forming disks \citep{MosqueiraPDS702026}.
Rafikov's stalling criterion distinguishes two branches: exterior clearing behind an
inward-moving body and migration feedback \citep{Rafikov2002}. The thin
protoplanetary cases of \citet{Li2009} select the latter branch. 
Our thicker satellite disk instead
selects the exterior-clearing threshold.

\citet{GinzburgSari2018} extended non-local wave deposition to deep gaps with
nonuniform gas density. Their equilibrium calculations show that, as gas near
the orbit is depleted, excitation farther from the planet becomes increasingly
important and can produce deeper, wider gaps in low-viscosity disks.
Finite disk thickness changes the torques acting on an embedded satellite
\citep{Tanaka2002,MenouGoodman2004,JimenezMasset2017} and therefore affects
the balance between migration and disk clearing. 

Local buoyancy deposition\footnote{Buoyancy-wave excitation and its implications
for angular-momentum transport and damping were already discussed by
\citet{MosqueiraEstrada2003b}.} provides an additional route for angular-momentum
transfer near the satellite's orbit, where non-local shock deposition leaves
residual gas in the single-satellite case \citep{MosqueiraBuoyancy2026}. In
multi-satellite systems, a satellite can redistribute the near-orbit gas of its companions.

Gap edges can be unstable to non-axisymmetric Rossby waves even when they
satisfy the Rayleigh criterion for stability. 
\citet{Lovelace1999} and \citet{Li2000} relate this
instability to extrema of an entropy-weighted inverse vortensity and show how
localized disk structure can support growing Rossby modes. Their nonlinear
development can produce vortices and additional angular-momentum transport.
Rayleigh adjustment alone therefore does not preclude the development of 
Rossby-wave instabilities at gap edges.

\citet{McNally2019} find that stochastic vortex
forcing and gas redistribution at
gap edges can sustain
inward migration in the single-planet case. 
In their inviscid giant-planet calculations, migration can
also stall after an inward displacement comparable to the gap width, as
redistribution of the outer pressure bump ceases to sustain inward drift.
Establishing sustained migration in planetary disks requires treatment 
of gas redistribution and torques in three-dimensional
disks with multiple-body effects. Planet-forming disks generally produce several
planets with overlapping growth profiles.

The same is true for satellite disks, where multiple satellites can form and
redistribute gas collectively. In addition, the significance of
stochastic-forcing migration remains to be assessed:
moderate viscosity can suppress its effects, its relevance
needs to be evaluated in 3D simulations, published work addresses
only the feedback branch of
Rafikov's stalling criterion, and its role in the multi-body context
is expected to be more limited.

Furthermore, extended disk clearing could make it
possible for later generations of satellites to survive even if the first
generation were lost. Indeed, the formation of Rossby-wave-unstable gap edges
implies a regime in which gas redistribution is already
significant. Finally, it is also important to note that
the observations of existing Solar System satellites suggest that 
we should expect both satellite loss and survival.\footnote{The same may not
be true for the satellite systems around PDS~70; see \citet{MosqueiraPDS702026}.}

This paper compares eight histories with the same initial disk
(Table~\ref{tab:case_map}), including matched Ganymede and Callisto-pair
calculations that suppress gap-edge pressure bumps while conserving mass and
angular momentum
(Section~\ref{sec:pressure_edges}).
Resonant configurations in satellite systems also motivate us to examine
whether collective disk clearing and migration slowing are compatible with
resonance capture.
Stochastic edge forcing for isolated and multiple satellites, self-consistent
eccentric disk--satellite coupling and its effects on resonant evolution, and
unresolved near-orbit buoyancy transport are reserved for a forthcoming paper.
Future work will also examine how inner-boundary reflection affects low-order
wave excitation and deposition, together with disk self-gravity and apsidal
precession from the disk's axisymmetric gravitational potential.

Section~\ref{sec:disk} specifies the disk;
Sections~\ref{sec:excitation}, \ref{sec:transport} and \ref{sec:numerics} describe
excitation, transport and numerical evolution. Section~\ref{sec:current_evolution}
presents the isolated Ganymede results; Section~\ref{sec:gradient_sensitivity}
gives the gradient-sensitivity and stress tests. Sections~\ref{sec:pair_extension} and
\ref{sec:buoyancy_summary} present the reference Callisto pair and buoyancy.
Section~\ref{sec:pressure_edges} examines clearing and resonant migration
with gap-edge redistribution.
Section~\ref{sec:conclusions} gives the conclusions.
Appendices~\ref{app:three_d}--\ref{app:rayleigh} provide the calibration, buoyancy
coupling, numerical specifications, deposition scalings and details of
adjustment that conserves mass and angular momentum.

\begin{table}[!htbp]
  \centering
  \small
  \renewcommand{\arraystretch}{0.9}
  \caption{Eight saved histories. Every isolated case begins with Ganymede at
  $20R_J$; the Callisto pair begins at $20$ and $25R_J$. All use the same initial
  disk and background viscosity. The radial-stability diagnostic is
  $\chi=\min(\kappa_3^2,\kappa_5^2)/\Omega_{\rm pb}^2$, where
  $\Omega_{\rm pb}$ is the pressure-balanced angular frequency and
  $\kappa_3^2$, $\kappa_5^2$ are three- and five-point estimates of the squared
  radial epicyclic frequency. Thus $\chi=1$ in the initial disk and $\chi=0$
  at Rayleigh marginality. The transport laws are specified in
  Sections~\ref{sec:gradient_sensitivity} and \ref{sec:pressure_edges}.}
  \label{tab:case_map}
  \begin{tabular}{p{0.22\textwidth}p{0.23\textwidth}p{0.34\textwidth}c}
    \toprule
    Case & Ordinary excitation & Disk redistribution & Buoyancy\\
    \midrule
    Ganymede baseline & Original modal & Instantaneous Rayleigh & No\\
    Calibrated Ganymede & Finite thickness & Instantaneous Rayleigh & No\\
    Buoyancy comparison & Same calibrated & Instantaneous Rayleigh & Yes\\
    Callisto pair & Original modal & Instantaneous Rayleigh & No\\
    Ganymede, pressure \mbox{adjustment} & Finite thickness & Rayleigh + pressure adjustment & No\\
    Callisto pair, pressure \mbox{adjustment} & Original modal & Rayleigh + pressure adjustment & No\\
    Gradient sensitivity & Same calibrated & Finite, onset $\chi=0.5$ & No\\
    Early-onset stress test & Same calibrated & Finite, onset $\chi=1$ & No\\
    \bottomrule
  \end{tabular}
\end{table}

\section{Disk and initial conditions}
\label{sec:disk}

\begin{figure}[!htbp]
  \centering
  \includegraphics[width=\textwidth]{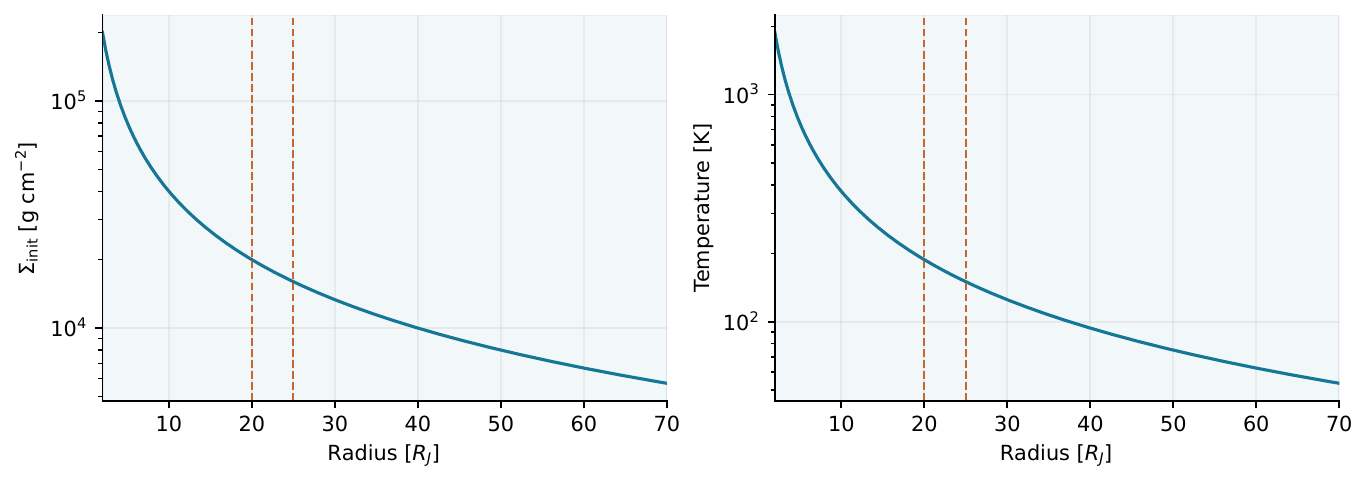}
  \caption{The actual initial disk used by all reported calculations: surface
  density (left) and prescribed temperature (right). Markers show the $20R_J$
  isolated-Ganymede orbit and the $20$, $25R_J$ Callisto-pair orbits. The shaded
  region is the modeled gas domain, $2$--$70R_J$. No radial transition is
  imposed.}
  \label{fig:disk}
\end{figure}

Let $R$ be cylindrical distance from Jupiter, $t$ elapsed time, $\Sigma(R,t)$
gas surface density and $T(R)$ the prescribed temperature. The initial surface
density $\Sigma_{\rm init}(R)=\Sigma(R,0)$ is specified by the residual-gas profile
\begin{equation}
  \Sigma_{\rm init}(R)=2\times10^4\left(\frac{R}{20R_J}\right)^{-1}
  {\rm g\,cm^{-2}},\qquad T(R)=\frac{3750}{R/R_J}\ {\rm K}.
  \label{eq:disk}
\end{equation}
Figure~\ref{fig:disk} shows the initial disk. This declining profile is an
idealized survival experiment motivated by the dense circumplanetary setting; it
is not a reconstruction of the original formation disk or its radial transition.
The formation inventory and the residual gas used for migration are distinct
\citep{MosqueiraEstrada2003a,MosqueiraEstrada2003b}. The gas normalization is
fixed before any evolution.

Reported physical values are rounded to two or three significant figures;
percentages and ratios are calculated before rounding the displayed values.
Appendix~\ref{app:reproduction} retains the numerical inputs used in the
calculations. Jupiter's mass and radius are $M_J=1.90\times10^{30}$ g and
$R_J=7.15\times10^9$ cm. We use adiabatic index $\gamma=1.4$ and molecular
mass $2.3m_H$, where $m_H$ is the hydrogen-atom mass. With gravitational
constant $G$, Boltzmann's constant $k_B$ and Keplerian angular frequency
$\Omega_K(R)=\sqrt{GM_J/R^3}$, the adiabatic sound speed $c_{\rm ad}$,
acoustic aspect ratio $h$ and background kinematic viscosity $\nu$ are
\begin{equation}
  c_{\rm ad}^2=\gamma k_BT/(2.3m_H),\quad h\equiv h_{\rm ad}=\frac{c_{\rm
  ad}}{R\Omega_K}
  =0.103,\quad
  \nu=\alpha_{\rm ad}h^2R^2\Omega_K,\quad \alpha_{\rm ad}=10^{-6}.
  \label{eq:viscosity}
\end{equation}
The dimensionless coefficient $\alpha_{\rm ad}$ uses the adiabatic sound speed;
unqualified $\alpha$ refers to this convention.
The disturbance coefficients assume an adiabatic response: thermal relaxation is
slow compared with the relevant forcing period. The mean temperature is
prescribed to isolate mechanical redistribution; dissipative heating is not fed
back into the disk structure. No radiative-transfer or thermal-energy evolution
is solved. Disk self-gravity is omitted.

The gas grid extends from $2$ to $70R_J$ with 801 uniformly spaced nodes. The
evolved interior gas inventory is $8.72\times10^{27}$ g. The inner boundary
allows drainage onto Jupiter without an imposed gas supply, with zero ordinary
viscous stress at its gas face; its endpoint density copies the adjacent
interior value. The outer endpoint retains $\Sigma_{\rm
init}(70R_J)=5.71\times10^{3}\ {\rm g\,cm^{-2}}$. Resonances beyond the outer
edge sample the prescribed unperturbed profile; resonances below $2R_J$ sample
zero gas. These boundary choices are fixed in all reported evolutions.

An isolated satellite of mass $M_G=1.48\times10^{26}$ g starts on a circular orbit
at $20R_J$. The pair has two masses $M_C=1.08\times10^{26}$ g, each $0.727M_G$,
initially circular at $20$ and $25R_J$ and separated by $180^\circ$ in longitude.
All reported evolutions start from equation~\eqref{eq:disk}, without an inherited
gap. The reference histories target 1,000 yr, subject to the recorded numerical
and orbital guards. The pressure-bump calculations have requested
computational endpoints near 400 and 472 yr, specified in
Appendix~\ref{app:reproduction}.

\section{Excitation and deposition in an evolving gap}
\label{sec:excitation}

\subsection{Spatially sampled modal torques}

For a satellite of mass $M_s$, semimajor axis $a$ and eccentricity $e$, define
$\mu=M_s/M_J$. Subscript $i$ identifies a satellite when more than one is
present. We calculate the positive magnitudes of the inner and outer launched
angular-momentum fluxes separately. The circular Lindblad reduction
uses the pressure cutoff of \citet{PapaloizouLarwood2000}. For azimuthal Fourier
mode number $m$, put $k_m=\sqrt{1+(mh)^2}$ and label the outer and inner branches with $\sigma=+1$
and $-1$. With dimensionless forcing coefficient $C_{m,\sigma}$, the launch
radii and positive flux amplitudes are
\begin{equation}
  R_{m,\sigma}=a(1+\sigma k_m/m)^{2/3},\qquad
  A_{m,\sigma}=C_{m,\sigma}(h)\Sigma(R_{m,\sigma})\frac{GM_s^2a}{M_J}.
  \label{eq:modal}
\end{equation}
We retain outer $m=1,\ldots,256$ and inner $m=2,\ldots,256$, with no finite
inner $m=1$ resonance or added continuum contribution. The outer $m=1$ potential
includes the planet-centered indirect term. Launch densities are linearly
interpolated on the evolving profile.

Appendix~\ref{app:modal_coefficient} gives the forcing coefficient and its
quadrature. Excitation uses Keplerian rotation with pressure-shifted
launch radii, without solving the full pressure-supported response in sharp gaps.
The two baseline evolutions use this two-dimensional reduction; the Ganymede
repeat applies the correction below. Ordinary corotation torque is zero under
assumed saturation. Dynamical corotation torques are not modeled; their absence
does not follow from ordinary saturation (Section~\ref{sec:buoyancy_summary}).

The satellite torque is
\begin{equation}
  \Gamma=A_--A_+,\qquad A_\pm=\sum_m A_{m,\pm}.
  \label{eq:net}
\end{equation}
The resonance densities, rather than the orbital column alone, determine
excitation strength and the inner--outer balance.

\subsection{A calibrated finite-thickness extension}
\label{sec:three_d}

Following \citet{MenouGoodman2004}, we soften the direct gravitational kernel with
$b_{\rm soft}(R)=\eta H_z(R)$, where $\eta$ is dimensionless,
$c_{\rm iso}=c_{\rm ad}/\sqrt\gamma$ is the isothermal sound speed and
$H_z=c_{\rm iso}/\Omega_K$ is the vertical density scale height. The squared
softened-to-original forcing ratio $\mathcal F_{m,\sigma}(\eta)$ multiplies each
amplitude, retaining launch radii, sampled density, pressure cutoff and the
indirect $m=1$ term. One coefficient thus supplies mode- and side-dependent
corrections while preserving excitation by distant gap-edge gas.

For the reference $\Sigma\propto R^{-1}$, $T\propto R^{-1}$ disk, $\eta=0.400$
matches the slow-diffusion Lindblad torque of \citet{JimenezMasset2017}. In units
$\Gamma_{0,\rm ad}=\Sigma(a)a^4\Omega_K^2(\mu/h)^2$, the inner/outer sums change
from $(8.95,14.4)$ to $(7.29,11.0)$, giving $\Gamma_L=-3.74\Gamma_{0,\rm ad}$:
total excitation falls by 21.6\% and net torque magnitude by 31.8\%.
Subscript $L$ denotes ordinary Lindblad torque.
Appendix~\ref{app:three_d} specifies the kernel, derivative, fit and spectral
data. The Ganymede repeat uses the corrected amplitude in both excitation and
shock onset (equation~\eqref{eq:3dshockchoice}); the original Ganymede and
Callisto histories retain equation~\eqref{eq:coefficient}.

\subsection{Launch-dependent shock estimate and retained tail}

For contributions allowed to damp, define the dimensionless launch offset
$x_0=|R_{m,\sigma}-a|/a$ and radial offset $x=|R-a|/a$. We use the shock-location
scaling of \citet{GinzburgSari2018}, equation 11,
\begin{equation}
  x_{\rm sh}=\max\left[x_0,
  \left(\frac{\Sigma_\sigma(x_{\rm sh})}{\Sigma_\sigma(x_0)}
  \frac{x_0^8h^3}{\mu^2}\right)^{1/5}\right],\qquad
  \Sigma_\sigma(x)=\Sigma(a+\sigma ax).
  \label{eq:shock}
\end{equation}
On each branch, we search from the launch point away from the satellite and
select the first admissible shock. The normalization is one, and no deposition
occurs between the launch point and the satellite. Applying this launch-annulus
scaling to individual harmonics, including $x_0<h$, is heuristic.

After this onset, an analytic surrogate with the Rafikov late-decay asymptote
spreads deposition downstream with a fixed width $w=0.264$, calibrated once on
a smooth, uniform-density
Ganymede control at $20R_J$. Grid-interval deposition follows by differencing its
cumulative fraction. Ginzburg \& Sari instead deposit at the shock without an
extended tail. Appendix~\ref{app:tail_definition} gives the full function and
calibration convention.

Spatially sampled excitation and downstream deposition permit tracking extended
gaps. Density sets launch strengths and shock locations, but the subsequent
geometric tail omits the evolving density along the post-shock path. It preserves
uniform-density homogeneity and a restricted local asymptotic scaling
(Appendix~\ref{app:deposition_density}), not damping through arbitrary structure.
Interference among contributions is omitted.

\subsection{The adopted low-mode escape limit}
\label{sec:escape}

The outer $m=1$ and both $m=2$ contributions retain their full excitation torques
but deposit no angular momentum within the modeled disk. Their complete signed
flux is counted as escaping at the corresponding boundary, with no reflection.
Higher modes retain the shock and tail prescriptions in
equations~\eqref{eq:shock} and~\eqref{eq:tail}; their undeposited
remainder also escapes.

\citet{GinzburgSari2018}, Section 5.1.2, caution that damping becomes uncertain
outside the local shearing-sheet limit, especially near $m\sim1$. Our $m\leq2$
cutoff retains distant excitation without extrapolated local clearing. It is an
assumed escape limit, not a derived threshold or fitted fraction; its coupled
orbital effect is calculated.

Partial reflection of inward low-order waves can weaken their secular torque
when damping is weak away from cavity resonances \citep{Tanaka2002,Tsang2011},
or enhance it through resonant standing waves \citep{Tsang2011,MirandaLai2018}.
Reflection does not guarantee shock dissipation: dispersion can oppose
steepening on the return path \citep{BrownOgilvie2026}. Including reflection
would therefore require a consistent treatment of both excitation and deposition.
We leave this extension to future work. The adopted draining condition for the
mean gas does not determine the wave boundary condition.

Here $j$ indexes a mode-and-side contribution, with positive launch magnitude
$A_j$ and branch sign $\sigma_j$. Let $P_j(R)$ be its deposition distribution
per unit radius and $q_j=\int P_j\,dR$ its fraction deposited in the modeled gas
domain. The signed deposited torque per unit radius $D(R)$ is positive when the
gas gains angular momentum, and $\dot J_{\rm esc}$ is the signed escape rate.
The accounting is
\begin{align}
  \Gamma&=-\sum_j\sigma_jA_j,&D(R)&=\sum_j\sigma_jA_jP_j(R),\nonumber\\
  \dot J_{\rm esc}&=\sum_j\sigma_jA_j(1-q_j),&
  \int D\,dR+\dot J_{\rm esc}&=-\Gamma.
  \label{eq:escape}
\end{align}
Every $m\leq2$ contribution has $P_j=q_j=0$. Higher-mode tails can have
$0<q_j<1$ on the finite domain. Inner escaping angular momentum is negative.
This secular accounting assumes negligible angular-momentum storage in waves.
The code evolves gas and orbits, not a wave field. Gas advection and viscous
boundary stresses are separate budget terms.

\section{Gas transport and Rayleigh adjustment conserving mass and angular momentum}
\label{sec:transport}

\subsection{Transport throughout gap formation}

Let $F_M$ be outward mass flux, $\ell_K=\sqrt{GM_JR}$ the specific Keplerian
angular momentum, and $G_\nu=3\pi\nu\Sigma\ell_K$ the positive outward viscous
torque. The retained mean-disk equations are
\begin{equation}
  2\pi R\partial_t\Sigma=-\partial_RF_M,\qquad
  F_M\frac{d\ell_K}{dR}=D-\partial_R(G_\nu+G_R),
  \label{eq:transport}
\end{equation}
where $G_R$ is the additional Rayleigh stress integrated as an outward torque;
both $G_R$ and $G_\nu$ have units of torque. The contributions from both
satellites enter the same $D$ and surface density. Background viscosity acts
during migration and clearing, not only as a test of whether an already formed gap
can be maintained. The ordinary wave and viscous fluxes use face-centered
$d\ell_K/dR$; their discrete error is included in the full budget residual.

\subsection{Instantaneous redistribution through a stress}

The baseline uses instantaneous adjustment, following the rapid-relaxation picture
of \citet{YangMenou2010}, implemented through fluxes conserving mass and Keplerian
angular momentum. The steady precedent of \citet{Kanagawa2015} instead uses
pressure-supported shear. Section~\ref{sec:gradient_sensitivity} replaces the
baseline constraint with finite local transport using the same flux construction.

Pressure-balanced rotation supplies the stability criterion, while transport
conserves the retained Keplerian inventory. Density and additional stress are
solved together at each secular step: stress vanishes in strictly stable regions
and permits redistribution at marginal stability. Compatible face fluxes cancel
interior transfers; the added stress vanishes at the boundaries while ordinary
drainage remains active. This is an instantaneous local adjustment, not a resolved
history of instability growth or a prediction of turbulent viscosity.
Appendix~\ref{app:rayleigh} specifies the stability diagnostic, stress conditions
and discrete conservation identity.

Adjustment also redistributes near-orbit gas where wave deposition is weak. Its
contribution to clearing and torque balance is recorded separately from wave and
background-viscous transport.

\section{Numerical evolution and diagnostics}
\label{sec:numerics}

\subsection{Gas and orbital integration}

We evolve the logarithm of the gas density implicitly to keep the density
positive, using backward Euler for the first step and variable-step BDF2
thereafter. Gas and orbital torques are coupled within each accepted step;
failed nonlinear steps are retried at smaller steps. No density floor, clipped
stress or fallback density reset is used. Cartesian orbits include mutual
gravity, the planet-centered indirect acceleration, disk torques and radial
eccentricity damping. There is no imposed resonant lock or stopping radius.
The pressure-adjusted cases additionally apply a mass- and angular-momentum-conserving
projection after the gas--orbit step. A nonzero adjustment resets the multistep
history, making this split extension first order in time.
Appendix~\ref{app:integration} gives the equations, step controls, solver
tolerances and guards; Appendix~\ref{app:pressure_edges} specifies the projection.

Eccentricity evolves under mutual gravity and disk forces, but excitation,
propagation and deposition retain the circular-wave approximation and linear
reference damping. The three baseline histories may continue beyond the diagnostic
scale $e=h$, which is not a physical instability boundary; their eccentric
portions do not establish long-term survival. The finite-transport and buoyancy
cases stop at $e\geq h$, though they and isolated 3D Ganymede remain far below it.
Self-consistent eccentric disk coupling is deferred
(Section~\ref{sec:introduction}). Numerical and orbital guards remain active throughout.

\subsection{Measured slowing and conservation}

We save checkpoints containing the disk, stress and Cartesian orbital state
approximately every two years, and orbital elements after each accepted gas
step. Reported fitted drift uses ordinary least squares over stated intervals; slowing
is $1-\dot a_{i,\rm fit}/\dot a_i(0)$ relative to the calculated initial inward
rate, allowing positive late slopes. These percentages are measured outputs.

The following decomposition concerns ordinary excitation; with buoyancy the
total orbital torque is $\Gamma_{\rm total}=\Gamma_L+\Gamma_B$, where $B$
denotes buoyancy. At each saved radius the initial profile in equation~\eqref{eq:disk} supplies a control under
the same modal prescription. Its signed torque $\Gamma_{L,0}(a)$ differs from
the positive normalization $\Gamma_{0,\rm ad}$ in Section~\ref{sec:three_d}.
Define $S_L=A_-+A_+$ and
$\epsilon_L=(A_+-A_-)/S_L$. Then
\begin{equation}
  \Gamma_L=-\epsilon_L S_L,\qquad
  \frac{\Gamma_L}{\Gamma_{L,0}(a)}
  =\frac{S_L}{S_{L,0}(a)}\frac{\epsilon_L}{\epsilon_{L,0}(a)}.
  \label{eq:decomposition}
\end{equation}
In ordinary-only diagnostics we omit the subscript $L$. The individual positive
amplitudes and signed torques distinguish diminished excitation from
cancellation. A nearly fixed semimajor axis need not mean zero angular-momentum
loss when eccentricity grows.

Separate ledgers record mass, added-transport angular momentum and the full
disk--orbit--boundary balance, including planetary reflex correction, boundary
advection, viscous stress and signed wave escape. No artificial Rayleigh source
is subtracted. Appendix~\ref{sec:conservation} reports both the added-flux
residual and remaining ordinary spatial/coupling errors. Positive density and
these budgets check the implementation. They do not establish mesh convergence;
all cases use one resolution.

\section{Isolated-satellite confinement and torque balance}
\label{sec:current_evolution}

We compare isolated Ganymede with original and calibrated excitation, then test
whether exterior depletion supplies the reversal. Gradient transport, buoyancy and
the eccentric pair are considered separately below.

\subsection{Rafikov's thresholds as initial-disk diagnostics}
\label{sec:rafikov}

\begin{figure}[!htbp]
  \centering
  \includegraphics[width=\textwidth]{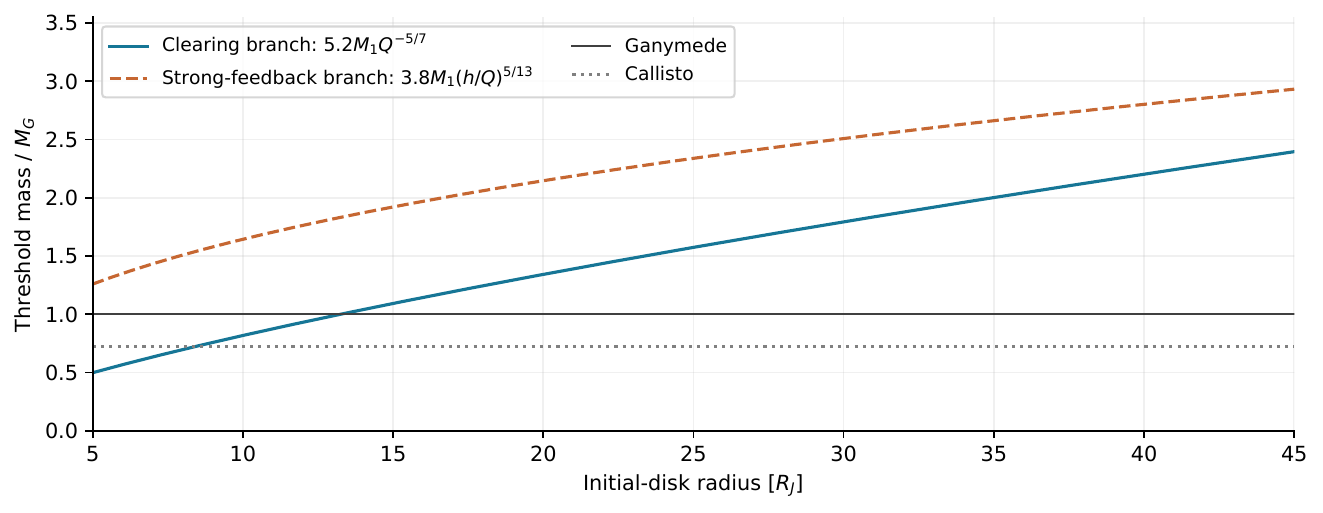}
  \caption{Published Rafikov thresholds evaluated on the actual initial $R^{-1}$
  gas disk, using its adiabatic sound speed. The clearing branch is lower than the
  strong-feedback branch. Horizontal lines indicate the two satellite masses.
  These curves are diagnostic only; the simulated gas and orbits determine the
  subsequent evolution.}
  \label{fig:rafikov}
\end{figure}

We compare the evolution with the published rounded thresholds of
\citet{Rafikov2002}, written in the familiar form quoted by \citet{Li2009}:
\begin{equation}
  M_{\rm clear}=5.2M_1Q^{-5/7},\qquad
  M_{\rm feedback}=3.8M_1(h/Q)^{5/13},\qquad
  M_1=\frac{2c_{\rm ad}^3}{3G\Omega_K},\quad
  Q=\frac{c_{\rm ad}\Omega_K}{\pi G\Sigma_{\rm init}}.
  \label{eq:rafikov}
\end{equation}
The ratio and the value of $Q$ at equality are
\begin{equation}
  \frac{M_{\rm clear}}{M_{\rm feedback}}=\frac{5.2}{3.8}h^{-5/13}Q^{-30/91},
  \qquad Q_\times=\left(\frac{5.2}{3.8}\right)^{91/30}h^{-7/6}.
  \label{eq:branch_selection}
\end{equation}
Thus $Q>Q_\times$ selects clearing, and the reverse inequality selects strong
feedback. For our $h=0.103$, $Q_\times=36.7$. The lower branch is the clearing
threshold over the radii shown in Figure~\ref{fig:rafikov}. On this specified
initial disk the Ganymede and Callisto mass lines cross that branch at
approximately $\RafG R_J$ and $\RafC R_J$, respectively. The calculation never
uses these crossings to stop a satellite. They are diagnostics of the unperturbed
disk, not predictions of the terminal radius after the gas has been redistributed.

Finite viscosity, discrete resonances, low-mode escape and Rayleigh transport
differ from equation~\eqref{eq:rafikov}'s assumptions. A separately preserved
historical inviscid calculation recovered 5.25 for the rounded clearing
coefficient 5.2; it does not validate this modal evolution. The formula supplies
the initial branch ordering and mass scale, not a solution of the original
translating-profile problem with our closure.

\subsection{An isolated Ganymede-mass satellite: the two-dimensional baseline}

The isolated body initially migrates inward at $\GInitialRate\,R_J\,{\rm
yr}^{-1}$. Clearing and redistribution alter that torque, and the orbit reaches a
minimum semimajor axis of $\GMinA R_J$, rebounds outward and enters a phase of
much smaller radial excursions (Figure~\ref{fig:3devolution}). At \GEnd{} yr the
orbit is $\GFinalA R_J$. The fitted final-100-year drift is $\GLastRate\,R_J\,{\rm
yr}^{-1}$, a \GLastReduction\% reduction from the initial inward rate. The maximum
eccentricity over the calculation is $\GMaxE$, so this case stays within the
small-eccentricity regime of the retained disk interaction.

The final gas column at the orbit is $\Sigma=\GFinalLocalSigma\ {\rm
g\,cm^{-2}}$, whereas the disk-wide minimum is $\GMinSigma\ {\rm g\,cm^{-2}}$.
Neither quantity alone specifies the resonance-weighted torque or its sign. The
reversal emerges without a stopping rule, with Rayleigh redistribution included
in the recorded budget.

The final same-radius ratios are $S/S_0=\GExcitationRatio$,
$\epsilon/\epsilon_0=\GBalanceRatio$ and $\Gamma/\Gamma_0=\GNetRatio$.
Equation~\eqref{eq:decomposition} separates reduced excitation from cancellation;
the small differential torque is not prescribed by scaling the orbital density.

\subsection{Ganymede with calibrated finite-thickness excitation}
\label{sec:three_d_evolution}

The repeated $20R_J$ Ganymede experiment fixes $\eta_\star=0.400$
(Appendix~\ref{app:three_d}) and corrects both excitation and shock amplitude.
Disk, physical mass, launch radii, tail, escape, viscosity, boundaries and
Rayleigh operator are unchanged; buoyancy is absent. Here ``3D'' denotes
calibrated excitation within the secular model, not a fluid simulation.

\begin{figure}[!htbp]
  \centering
  \includegraphics[width=\textwidth]{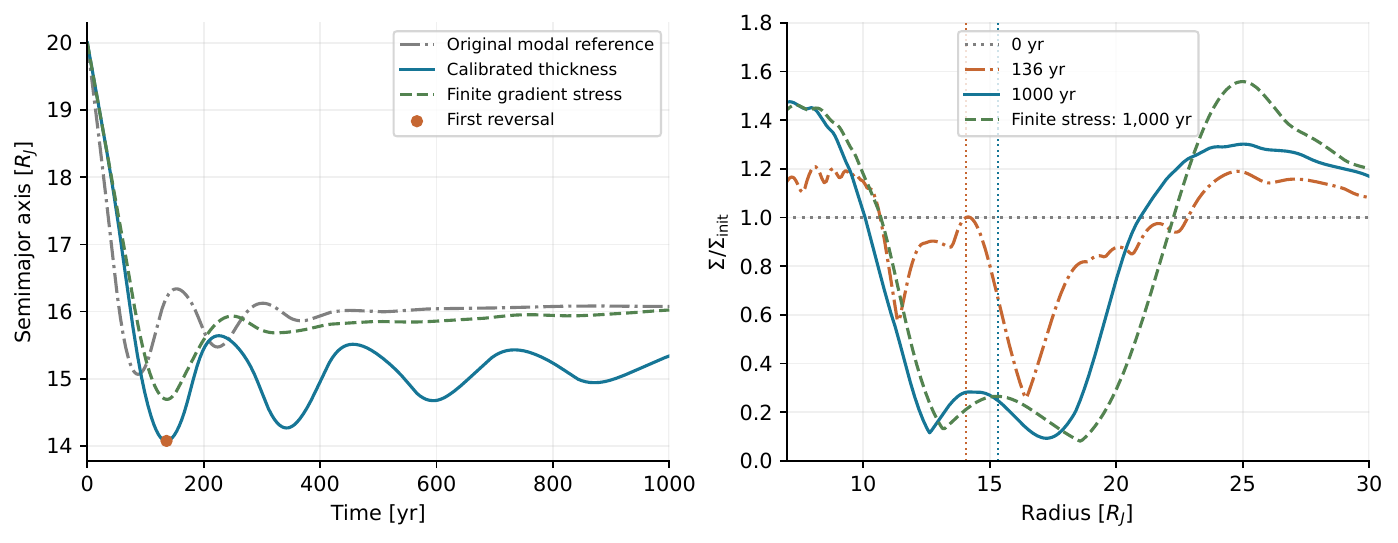}
  \caption{Completed isolated-Ganymede comparison. Left: the original modal
  reduction, calibrated finite-thickness evolution, and $\chi=0.5$
  finite-gradient-stress comparison (Section~\ref{sec:gradient_sensitivity}), all
  starting at $20R_J$. The point marks the first saved nonnegative torque in the
  calibrated case with instantaneous adjustment. Right: its gas profiles at
  initialization, the first saved reversal and 1,000 yr, divided by the initial
  profile. Dotted lines mark the corresponding evolved orbits. The dashed profile
  is the final finite-stress comparison. Outer depletion is already substantial at
  the first reversal, while gas remains near the satellite.}
  \label{fig:3devolution}
\end{figure}

The calculation completes 1,000 yr in 10,000 accepted steps with no rejected steps
(Figure~\ref{fig:3devolution}). Its initial inward rate is $-0.0492\,R_J\,{\rm
yr}^{-1}$. The first orbital minimum is $14.1R_J$ at 136 yr, followed by outward
motion and repeated radial excursions. The final orbit is $15.3R_J$; over the last
500 yr it remains between $14.7$ and $15.4R_J$. The least-squares drift over
500--1,000 yr is $+\CalibratedLateRate\,R_J\,{\rm yr}^{-1}$. The 800--900 and
900--1,000 yr slopes have opposite signs, $-2.79\times10^{-3}$ and
$+3.70\times10^{-3}\,R_J\,{\rm yr}^{-1}$. Thus the result is prolonged radial
confinement with oscillatory migration, rather than an exact fixed-radius
equilibrium at the endpoint. The maximum eccentricity is $2.34\times10^{-5}$.

At 1,000 yr the gas column at the orbit is $6.43\times10^3\ {\rm g\,cm^{-2}}$, or
0.246 of the initial same-radius column. The maximum recorded mass residual is
$5.74\times10^{-11}$ of the initial gas mass; the added-Rayleigh angular-momentum
residual is $3.94\times10^{-11}$ of the initial disk inventory. The full signed
endpoint residual is $-0.145\%$ of the absolute orbital angular-momentum change,
using the budget definitions in Appendix~\ref{sec:conservation}. The analysis
uses 498 disk profiles and 10,001 orbital states.

\subsection{Outer depletion is sufficient for torque cancellation}
\label{sec:outer_clearing}

To identify which density change supplies the first reversal, we reevaluate
every modal amplitude on the saved profile and on the initial disk at the
\emph{same} satellite radius. We use the same coefficients, launch radii,
satellite mass and interpolation rules in both evaluations. Write
\begin{equation}
  A_{\pm,0}(a)=\sum_m W_{m,\pm}(a)\Sigma_{\rm init}(R_{m,\pm}),\qquad
  W_{m,\pm}=\mathcal F_{m,\pm}C_{m,\pm}\frac{GM_G^2a}{M_J}.
  \label{eq:control_launch}
\end{equation}
Subscript 0 denotes the initial profile at current resonances, not the initial
$20R_J$ torque or an evolved viscosity-only control. All 511 contributions,
including escaping modes, enter.

An outer-depletion-only control retains the initial interior gas and admits only
density deficits at exterior resonances:
\begin{align}
  \Delta A_{+,\rm def}&=\sum_m W_{m,+}
  [\Sigma_{\rm init}(R_{m,+})-\Sigma(R_{m,+})]_+,\nonumber\\
  A_{+,\rm def}&=A_{+,0}-\Delta A_{+,\rm def},&
  \Gamma_{\rm outer\ only}&=A_{-,0}-A_{+,\rm def},
  \label{eq:outer_only}
\end{align}
where $[u]_+=\max(u,0)$. This excludes assistance from an inner pile-up and also
excludes any positive density excess on the outer side. The fractional outer
reduction needed to cancel the undepleted inward torque and the measured deficit
are
\begin{equation}
  \delta_{\rm req}=1-\frac{A_{-,0}}{A_{+,0}},\qquad
  \delta_{\rm def}=\frac{\Delta A_{+,\rm def}}{A_{+,0}}.
  \label{eq:outer_sufficiency}
\end{equation}
For inward control torque, $\delta_{\rm def}\geq\delta_{\rm req}$ is an exact
sufficiency test within this prescription, weighted over all resonances rather
than inferred from a density minimum or uniform reduction.

The total change also has the direct signed decomposition
\begin{equation}
  \Gamma-\Gamma_0=(A_{+,0}-A_+)-(A_{-,0}-A_-).
  \label{eq:direct_change}
\end{equation}
A positive exterior deficit weakens the inward torque, whereas a positive interior
deficit weakens the outward torque; density excesses reverse the corresponding
signs.

At the first saved nonnegative torque ($t=136$ yr, $a=14.1R_J$), exterior
depletion reduces the outer torque by 42.0\%, exceeding the required 33.9\%
(Table~\ref{tab:3dstall}). With no exterior excess at any launch, $A_{+,\rm
def}=A_+$. Restoring the initial interior gives $\Gamma_{\rm outer\
only}=+1.86\times10^{32}\ {\rm dyn\,cm}$; the actual inner amplitude has fallen
12.0\%, leaving $+3.43\times10^{30}\ {\rm dyn\,cm}$. Exterior loss supplies
124\% of the initial net inward torque, offset by 23.5\% from inner weakening.
An inner pile-up is unnecessary and does not cause this reversal.

\begin{table}[htbp]
\centering\small
\caption{Same-radius torque budget at the first saved 3D reversal ($t=136$ yr, $a=14.1R_J$). Inner and outer entries are positive magnitudes; torques are in $10^{33}\ {\rm dyn\,cm}$. The outer-depletion-only row retains the initial inner disk and only density deficits at outer resonances. Net torques are calculated before rounding the amplitudes.}
\label{tab:3dstall}
\begin{tabular}{lrrr}\toprule
Profile & $A_-$ & $A_+$ & $\Gamma=A_--A_+$\\\midrule
Undepleted control & 1.52 & 2.29 & -0.778 \\
Outer depletion only & 1.52 & 1.33 & +0.186 \\
Actual evolving profile & 1.33 & 1.33 & +0.00343 \\
\bottomrule\end{tabular}\end{table}

\begin{figure}[!htbp]
  \centering
  \includegraphics[width=\textwidth]{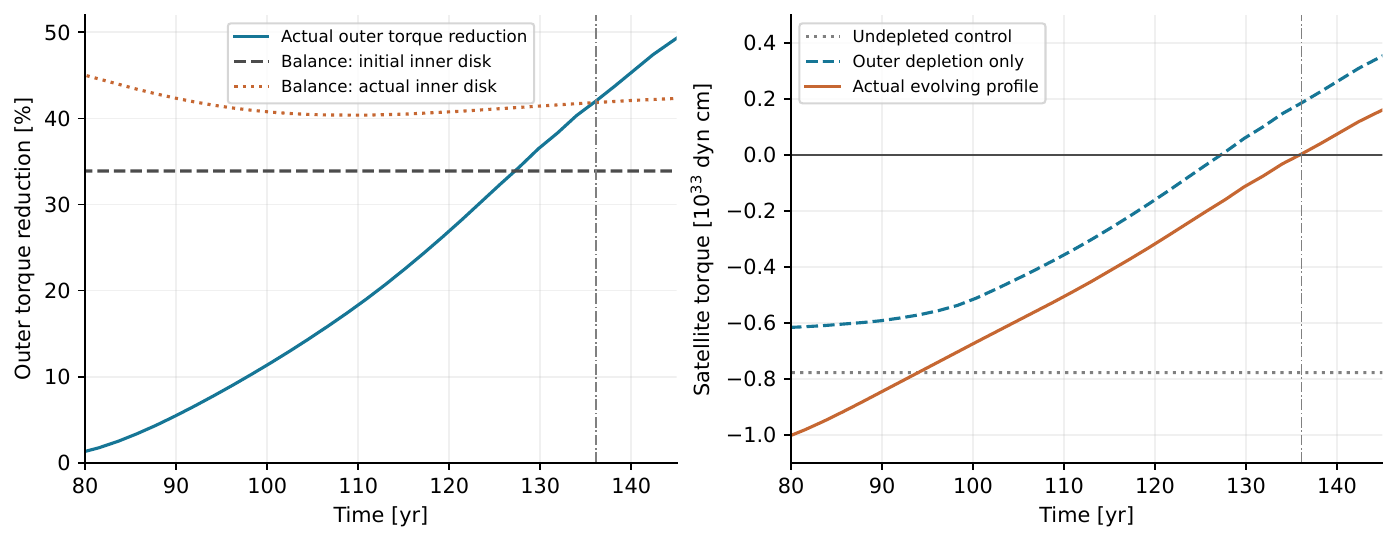}
  \caption{Outer-clearing sufficiency test during the first 3D reversal. Left: the
  actual outer torque reduction, including any exterior density excess, exceeds
  the reduction required to balance the initial inner disk before reaching the
  larger reduction required by the actual, weakened inner torque. Right: signed
  torques at the current satellite radius; the outer-depletion-only control
  retains deficits but excludes density excesses. The vertical line marks the
  first saved nonnegative actual torque, at 136 yr, when no outer excess remains
  at the launch sites. The outer-depletion-only control reverses earlier and has a
  larger outward torque. No percentage is an imposed suppression factor.}
  \label{fig:3dclearing}
\end{figure}

Figure~\ref{fig:3dclearing} follows this test through the approach to the first
stall. The outer-depletion-only torque reverses near 127 yr at $14.1R_J$, before
the actual reversal near 136 yr. The independently saved 0.1-yr orbital history
also places the first minimum near 136 yr. Appendix~\ref{app:reproduction} gives
the precise saved-time brackets and interpolation values, distinguishing the
resolution of this diagnostic from that of the orbital integration. These control
profiles are evaluated on saved states; they are not separate histories with
feedback suppressed.

At $a=14.1R_J$, initial-disk $Q=216$ gives $M_{\rm clear}=1.04M_G$ and $M_{\rm
feedback}=1.88M_G$. The first turn is about 6\% outside the $13.2R_J$ clearing
crossing. The lower clearing threshold and sufficient outer torque reduction
support clearing near Rafikov's predicted mass and radius scales. This
comparison does not reproduce his mathematical branch problem. Both sides
evolve, and Rafikov's general feedback includes outer depletion; our test
identifies exterior excitation loss, rather than enhanced inner excitation, as
sufficient here.

\section{Sensitivity to the treatment of disk gradients}
\label{sec:gradient_sensitivity}

We replace instantaneous adjustment with finite turbulent stress beginning while
the disk is Rayleigh stable. This tests earlier smoothing in the direction
motivated by \citet{McNally2019}'s gap-edge instabilities. The run starts with
Ganymede on a circular orbit at $20R_J$, retaining the calibrated Ganymede setup
of Section~\ref{sec:three_d_evolution} except for the added transport law below.

Let $\Omega_{\rm pb}$ be the pressure-balanced angular frequency, and
$\kappa_3^2$ and $\kappa_5^2$ the three- and five-point estimates of the squared
radial epicyclic frequency $\kappa_{\rm pb}^2$
(Appendix~\ref{app:rayleigh}). Define the normalized stability diagnostic
$\chi$ and additional dimensionless viscosity $\alpha_{\rm extra}$ by
\begin{equation}
  \chi=\frac{\min(\kappa_3^2,\kappa_5^2)}{\Omega_{\rm pb}^2},\qquad
  \alpha_{\rm extra}=10^{-3}\min\!\left[1,\max\!\left(0,
  \frac{0.5-\chi}{0.5}\right)\right].
  \label{eq:gradient_alpha}
\end{equation}
Transport begins at $\chi=0.5>0$, rising linearly to $10^{-3}$ at Rayleigh
marginality ($\chi=0$) and remaining capped below it. With $\nu_{\rm
extra}=\alpha_{\rm extra}c_{\rm ad}^2/\Omega_K$ and $G_{\rm extra}=3\pi\nu_{\rm
extra}\Sigma\ell_K$, this stress replaces $G_R$ in equation~\eqref{eq:transport},
using equation~\eqref{eq:sharedflux} and the same zero-stress boundary-adjacent
cells. The implicit step evaluates density and stress together, without projection
or a lower bound on $\chi$. Onset is phenomenological, not a calculated
Rossby-wave threshold. This axisymmetric redistribution test excludes stochastic
vortex forcing (Section~\ref{sec:introduction}).

The run completes 1,000 yr; its orbit and final profile are included in
Figure~\ref{fig:3devolution}. Ganymede reaches a minimum radius of $14.7R_J$ at
137 yr, reverses, and ends at $16.0R_J$. For comparison, instantaneous adjustment
gives a minimum of $14.1R_J$ at 136 yr and an endpoint of $15.3R_J$
(Section~\ref{sec:three_d_evolution}). Over 500--1,000 yr the finite-transport
orbit remains between $15.8$ and $16.0R_J$, compared with $14.7$--$15.4R_J$ for
instantaneous adjustment. The fitted drifts over this interval are
$+\GradientLateRate$ and $+\CalibratedLateRate\,R_J\,{\rm yr}^{-1}$, respectively.
Both prescriptions therefore yield prolonged radial confinement with small mean
outward motion; the detailed excursions differ.

At 1,000 yr the orbital column is $6.16\times10^3\ {\rm g\,cm^{-2}}$, or 0.247
of its initial same-radius value, versus 0.246 for instantaneous adjustment. The
minimum exterior density fractions are 0.0809 at $18.6R_J$ and 0.0928 at
$17.2R_J$, respectively, corresponding to about 92\% and 91\% depletion. Both
have maximum eccentricity $2.34\times10^{-5}$. The modest displacement is
outward; both remain confined near $15$--$16R_J$ with nearly identical local
depletion.

The largest recorded additional $\alpha$ is $6.24\times10^{-6}$, or 0.62\% of
its $10^{-3}$ cap, giving total local $\alpha=7.24\times10^{-6}$. The minimum
$\chi=0.497$ remains near onset; no saved state reaches the cap. This
prescription changes both when transport begins and how it responds to the
gradient, rather than changing the relaxation time alone. Depletion and
confinement persist. Maximum mass and added-angular-momentum residuals are
$1.39\times10^{-11}$ and $9.71\times10^{-12}$ of their initial disk inventories;
the full signed endpoint residual is $-0.0957\%$ of absolute orbital
angular-momentum change. Appendix~\ref{app:reproduction} documents the inputs
and diagnostics.

\subsection{Early-onset transport stress test and an inherited cavity}
\label{sec:early_transport}

The $\chi=0.5$ test already precedes Rayleigh instability and preserves stalling.
We push onset further by replacing $(0.5-\chi)/0.5$ with $(1-\chi)/0.5$ in
equation~\eqref{eq:gradient_alpha}: transport begins at $\chi=1$ and reaches the
same $10^{-3}$ cap at $\chi=0.5$, with unchanged clipping. The initial $R^{-1}$
profile has $\Omega_{\rm pb}^2/\Omega_K^2=\kappa_{\rm pb}^2/\Omega_K^2=1-2b$
with $b=h^2/\gamma$ (equation~\eqref{eq:stability}), hence $\chi=1$. The law
responds to departures
from that initial curvature instead of waiting halfway to marginality. Everything
else, including the buoyancy-free $20R_J$ initial state and numerical guards, is
unchanged. This stress test pushes the onset to the initial disk's stability
level; the realized stress remains below the cap.

Although inward migration progressively slows, this treatment prevents an
effective stall before Ganymede reaches the retained
$3.5R_J$ orbital guard at 935 yr. The instantaneous endpoint drift is
$-3.67\times10^{-3}\,R_J\,{\rm yr}^{-1}$; the common-cadence least-squares drift
over the final 100 yr, 835--935 yr, is $-4.29\times10^{-3}\,R_J\,{\rm yr}^{-1}$.
The calculation stops at the modeled orbital limit; it neither fails numerically
nor demonstrates an impact on Jupiter. The maximum eccentricity is
$2.36\times10^{-5}$. The largest recorded additional $\alpha$ is
$2.66\times10^{-5}$ (total local $\alpha=2.76\times10^{-5}$), only 2.66\% of its
cap, and the minimum recorded $\chi$ is 0.987. Thus the changed outcome occurs
while the disk remains close to the much earlier activation threshold, without
realizing $\alpha_{\rm extra}=10^{-3}$.

The coupled satellite--disk evolution nevertheless leaves a broad depleted region.
At the endpoint the column at Ganymede is $7.8\times10^3\ {\rm g\,cm^{-2}}$, or
6.85\% of the initial column at the same radius. From the inner gas face at
$2.04R_J$, the remaining column is below 10\% out to $4.05R_J$, below 50\% out to
$9.46R_J$, and below 90\% out to $18.7R_J$. These radii interpolate threshold
crossings in the saved ratio $\Sigma(R,t)/\Sigma(R,0)$. The distinct exterior
trough at approximately 400 yr lies at $10.6R_J$ and retains 58.6\% of its initial
column. It subsequently merges into the inner cavity: at the endpoint there is no
separate exterior local minimum, and the first exterior sampled node, $3.53R_J$,
retains 7.00\%. The absolute density minimum, $5.23\times10^3\ {\rm g\,cm^{-2}}$
at $2.68R_J$, is near the draining boundary and is not an exterior trough.

Preventing the first stall therefore does not prevent substantial clearing. A
subsequent satellite encounters this inherited depletion; further orbital
angular-momentum loss passes to the remaining gas and boundary fluxes, while
displacement away from effective resonances further weakens the torque. On these
angular-momentum budget grounds we expect the next satellite to stall.

\section{Two Callisto masses: shared clearing and stalling}
\label{sec:pair_extension}

\begin{figure}[!htbp]
  \centering
  \includegraphics[width=\textwidth]{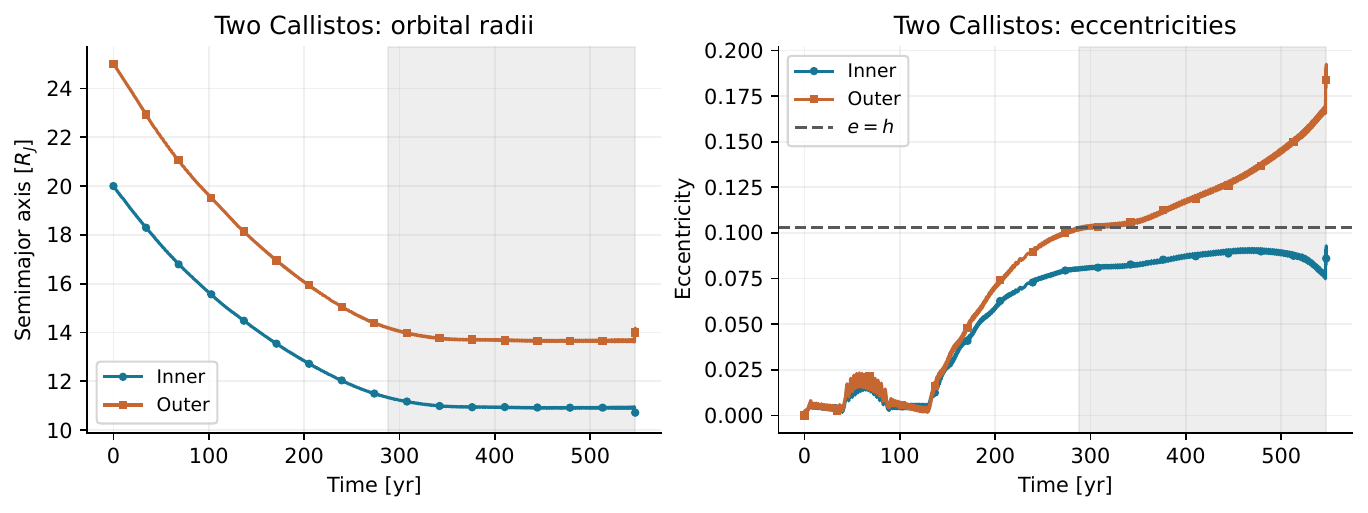}
  \caption{The Callisto pair migrates inward, stalls, and then undergoes a late
  encounter. Left: semimajor axes from every accepted gas step. Right:
  eccentricities. Gray shading begins at the outer body's first $e=h$ crossing,
  288 yr (dashed line). The 400--500 yr stalled interval precedes the final
  encounter-related excursion and integrator guard. Shaded intervals retain
  circular-wave disk coupling.}
  \label{fig:orbits}
  \medskip
  \centering
  \includegraphics[width=\textwidth]{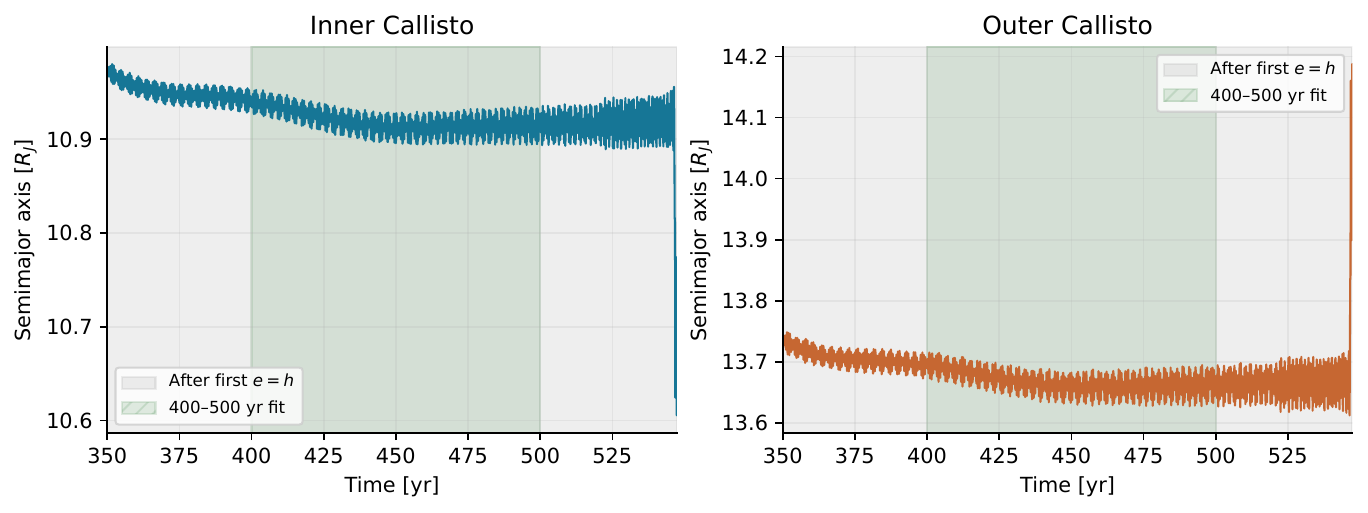}
  \caption{The stalled pair before its late encounter. The hatched 400--500 yr
  interval gives approximately 99.5\% slowing; the 450--540 yr fitted slopes are
  slightly outward. Small oscillations precede the final encounter-related
  excursion. Gray shading marks continued circular-wave coupling after $e=h$.}
  \label{fig:pairlate}
\end{figure}

The two Callisto masses clear a shared depleted region and settle into nearly
steady orbits before a late encounter (Figures~\ref{fig:orbits} and
\ref{fig:pairlate}). Their initial inner/outer migration rates are $-0.0524$ and
$-0.0586\,R_J\,{\rm yr}^{-1}$. The 180--280 yr fits show 63.5\% and 59.1\%
slowing; at 400--500 yr both have slowed by approximately 99.5\%, near $10.9$
and $13.7R_J$. The 450--540 yr fitted slopes are slightly outward
(Table~\ref{tab:pair_regime}). Stalling therefore precedes the encounter.

The eccentricity approximation is a separate limitation. The outer body first
crosses $e=h$ near 288 yr, before both stalled intervals; the precise bracket is
in Appendix~\ref{app:reproduction}. The 180--280 yr fit uses the 100-yr window
ending at the last multiple of 10 yr before crossing. This interval was not
selected to optimize the fitted slope. Its near-threshold eccentricities do not
validate the small-eccentricity approximation. The outer body's maximum $e/h$
reaches 1.42 over 400--500 yr and 1.61 over 450--540 yr. Circular excitation,
deposition and reference damping are retained, while mutual gravity is
integrated. The orbit and torque figures shade the post-crossing continuation;
Figure~\ref{fig:density} identifies its profile curves. These limits qualify the
eccentric disk response, not the recorded sequence of stalling followed by an
encounter.

\begin{table}[htbp]\centering\small
\caption{Pair drift and eccentricity in specified intervals. Body order is inner, outer. The first outer-body crossing of $e=h$ is near 288 yr; its precise bracket is in Appendix~\ref{app:reproduction}. The two later intervals measure the established stalled phase before the late encounter, using the continued circular-wave closure. Remaining drift is $100\dot a/\dot a(0)$; negative values denote outward motion relative to the initial inward drift.}
\label{tab:pair_regime}
\begin{tabular}{lrrr}\toprule
Interval [yr] & $\dot a_1,\dot a_2$ [$R_J\,{\rm yr}^{-1}$] & Remaining drift [\%] & $\max(e_1/h),\max(e_2/h)$\\\midrule
180--280 & $-1.91\times10^{-2},\ -2.40\times10^{-2}$ & 36.5, 40.9 & 0.782, 0.988\\
400--500 & $-2.37\times10^{-4},\ -3.07\times10^{-4}$ & 0.452, 0.523 & 0.888, 1.42\\
450--540 & $+7.64\times10^{-5},\ +8.33\times10^{-5}$ & -0.146, -0.142 & 0.888, 1.61\\
\bottomrule\end{tabular}
\end{table}

The late encounter follows the stalled phase and produces the final orbital
excursion. The encounter/integrator guard ends the run at \PairEnd{} yr, with
accepted semimajor axes $10.6$ and $14.1R_J$ and eccentricities 0.0912 and 0.190. The
rejected trial establishes no collision; neither body reaches the inner orbital
boundary. Mutual gravity is always integrated, but a period ratio alone does not
establish resonance. Eccentric disk coupling is deferred
(Section~\ref{sec:introduction}). Without an isolated Callisto control, shared
clearing does not establish a cooperative reduction of survival mass.

\subsection{Density depletion and one-sided excitation}

\begin{figure}[!htbp]
  \centering
  \includegraphics[width=0.96\textwidth]{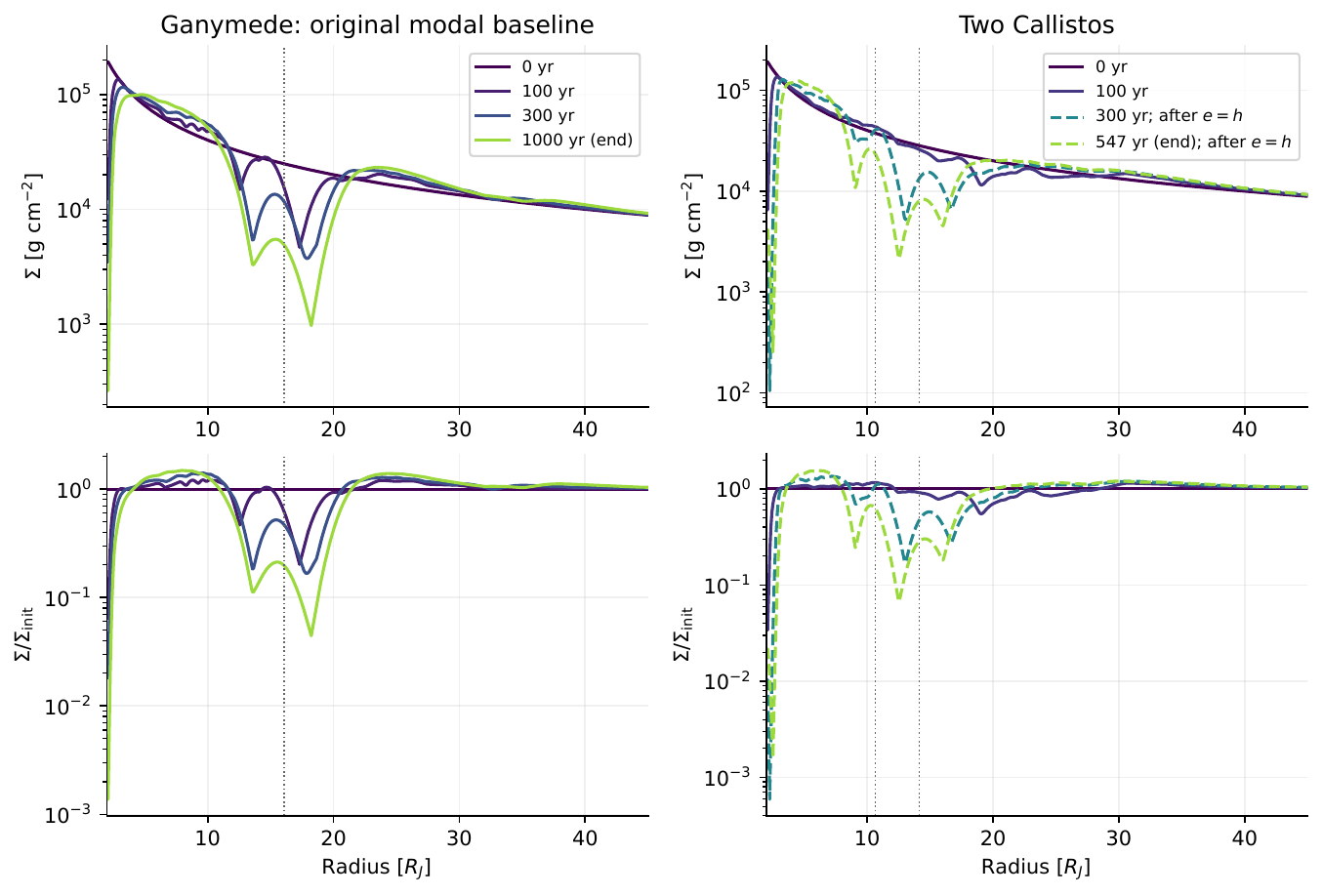}
  \caption{Gas profiles at selected saved times. Left: Ganymede, original modal
  baseline. Right: two Callistos. Absolute surface density is above and its ratio
  to the initial profile below. Dotted vertical lines mark the last accepted
  semimajor axes. Dashed pair-profile curves denote saved times after the first
  $e=h$ crossing; temporal shading is not placed on this radial axis. The initial
  profile is a reference, not a viscosity-only control. Both deposition and
  Rayleigh transport that conserves mass and angular momentum contribute to the redistribution shown. No
  density floor is applied.}
  \label{fig:density}
  \medskip
  \centering
  \includegraphics[width=0.96\textwidth]{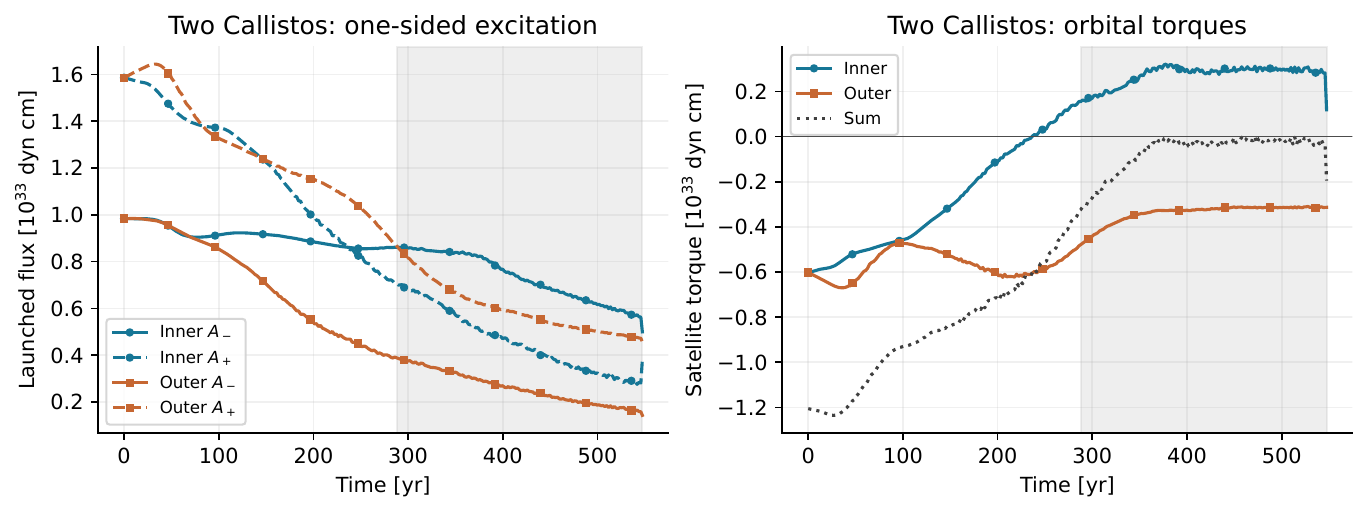}
  \caption{Pair excitation and orbital torques. Left: positive inner and outer
  launched angular-momentum fluxes $A_-$ and $A_+$ (solid and dashed lines).
  Right: signed satellite torques. Circle and square markers distinguish the
  inner and outer bodies, respectively; their summed torque is dotted.
  Gray shading marks its continuation after the first $e=h$ crossing. All modes
  contribute to excitation, including $m\leq2$, whose flux is not deposited in the
  modeled gas. The balance emerges from density sampled at the evolving
  resonances.}
  \label{fig:torques}
\end{figure}

Figure~\ref{fig:density} follows the gas redistribution. The pair excavates a
trough between the moons while the gas near each orbit can retain a substantially
larger column. At the last accepted pair state the minimum between the semimajor
axes is $2.14\times10^3\ {\rm g\,cm^{-2}}$ at $12.5R_J$, or $\PairGapRatio$ of the
initial column at that radius. The local orbital columns of the inner
and outer bodies are approximately $2.35\times10^4$ and $8.02\times10^3\ {\rm
g\,cm^{-2}}$, respectively 0.626 and 0.283 of their initial same-radius values.
The lower global minimum, $254\ {\rm g\,cm^{-2}}$, is not the inter-satellite
trough. Thus neither the global minimum nor the local orbital density alone
describes the torque-producing gas.

The independently summed launch amplitudes and signed torques of the pair are
shown in Figure~\ref{fig:torques}.

The satellites exchange angular momentum with each other, so their individual
disk torques need not vanish during the stall. Their final values,
$+1.18\times10^{32}$ and $-3.14\times10^{32}\ {\rm dyn\,cm}$, describe the late
eccentric encounter, following the stalled phase, rather than a circular
equilibrium or the earlier stalled balance. The outer body's eccentricity also
exceeds $e=h$ during the stalled interval.

\section{Buoyancy, local clearing and a smoother stall}
\label{sec:buoyancy_summary}

Ordinary deposition away from the orbit leaves nearby gas while adjacent regions
clear, permitting radial excursions (Figure~\ref{fig:3devolution}). Buoyancy adds
a near-orbit interaction under the assumed unsaturated, slow-relaxation limit
\citep{ZhuStoneRafikov2012,LubowZhu2014}.

Three-dimensional calculations also find that buoyancy changes the vortensity
of librating coorbital gas, producing a negative dynamical corotation torque
\citep{McNally2020,Ziampras2024}. That response is distinct from the direct
buoyancy exchange included here. We do not evolve coorbital vortensity or include
its dynamical torque; ordinary saturation does not eliminate this contribution.
The smoother confinement below therefore describes the specified local-deposition
comparison. Its extension to include dynamical corotation in a depleted satellite
disk remains future work.

Unlike the retained-heating analysis of \citet{MosqueiraBuoyancy2026}, this
mechanical comparison fixes temperature. A fresh Ganymede starts at $20R_J$ with
the calibrated Ganymede setup of Section~\ref{sec:three_d_evolution}, adding only
the buoyancy source below. No thermal slowing coefficient is imported.

With $D_{L,i}$ the ordinary deposition from satellite $i$ and
$\mu_i=M_i/M_J$, we add the signed buoyancy source $D_{B,i}$ using the kernel
$\mathcal K_{B,i}$ specified in Appendix~\ref{app:buoyancy_coupling}:
\begin{equation}
  D_{B,i}=\mu_i^2\Sigma(R,t)\mathcal K_{B,i}(R;a_i),\qquad
  D_{\rm total}=\sum_i(D_{L,i}+D_{B,i}),\qquad
  \Gamma_{B,i}=-\int D_{B,i}\,dR.
  \label{eq:buoyancycomparison}
\end{equation}
The kernel samples evolving buoyancy-resonance densities with explicit vertical
structure and no second acoustic thickness correction. Its signed spatial
integrals supply equal and opposite angular-momentum exchanges to the gas and
satellite, using matching time weights (Appendix~\ref{app:buoyancy_coupling}).
Initially, inner/outer gas torques are $-6.94\times10^{32}$ and
$+7.09\times10^{32}\ {\rm dyn\,cm}$: reaction $-1.54\times10^{31}\ {\rm
dyn\,cm}$, or 1.98\% of ordinary inward torque. Their magnitude sum,
$1.40\times10^{33}\ {\rm dyn\,cm}$, is 36.8\% of the ordinary sum
$3.81\times10^{33}\ {\rm dyn\,cm}$ (26.9\% of the combined sum). Small net
migration torque thus coexists with substantial local exchange.

\begin{figure}[!htbp]
  \centering
  \includegraphics[width=\textwidth]{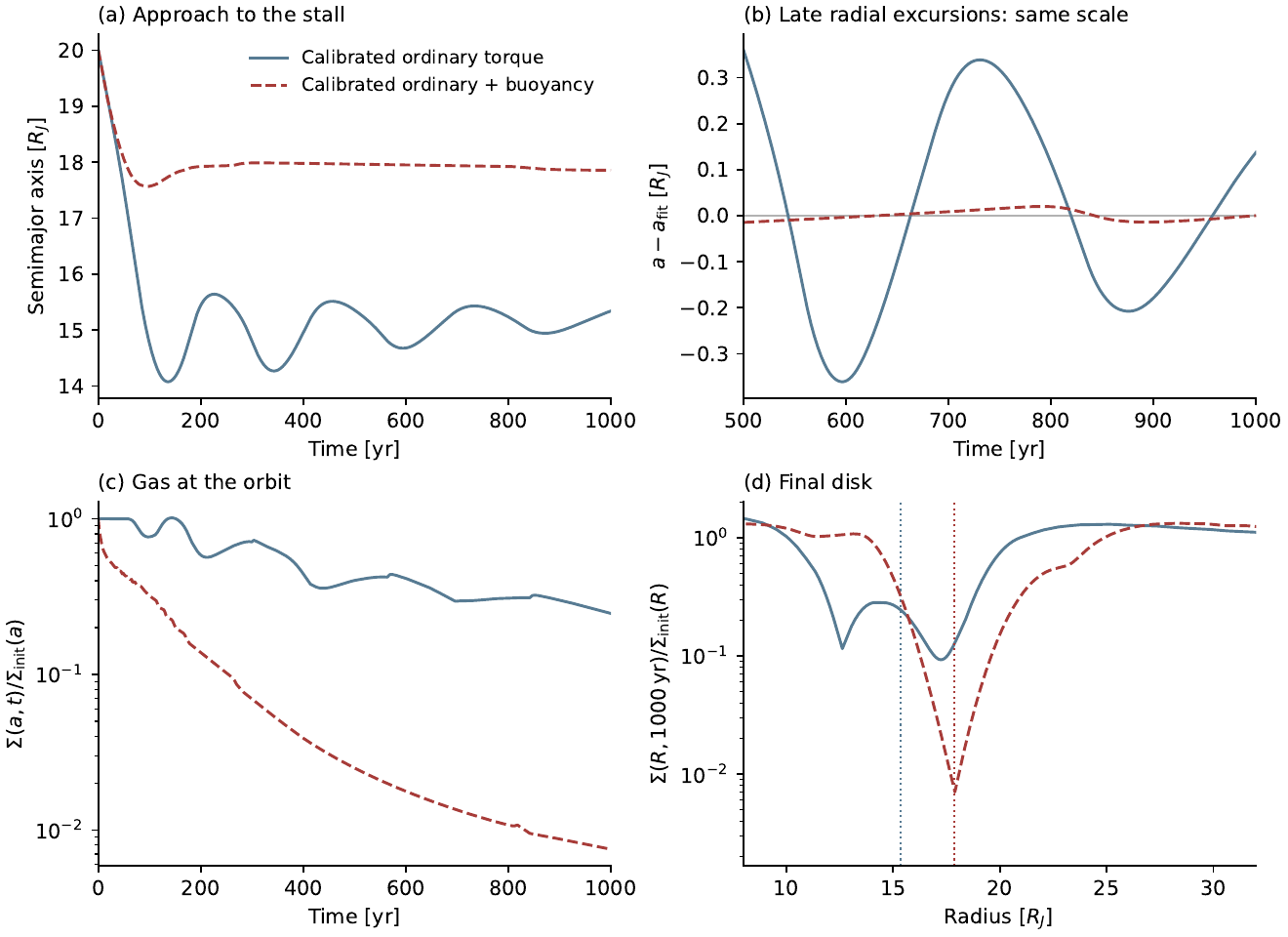}
  \caption{Matched isolated-Ganymede evolutions with the same ordinary 3D
  excitation and instantaneous Rayleigh transport conserving mass and angular
  momentum, without and with
  local buoyancy deposition. (a) Full orbital approach to the effective stall. (b)
  The common 500--1,000 yr interval after subtracting a separate least-squares
  linear trend $a_{\rm fit}(t)$ from each orbit, plotted on the same radial
  scale; buoyancy
  substantially reduces the remaining excursions. (c) Gas at the instantaneous
  orbit, divided by the initial column at that radius. (d) Final gas profiles
  relative to the initial disk; dotted lines mark the final satellite radii.
  Temperature remains fixed and buoyancy's orbital reaction is paired with its gas
  exchange.}
  \label{fig:buoyancyevolution}
\end{figure}

The calculation with buoyancy completes 1,000 yr and ends at $\BuoyFinalA R_J$,
with fitted 500--1,000 yr drift $\BuoyLateRate\,R_J\,{\rm yr}^{-1}$. The drift
magnitude is $\BuoyRateReduction\%$ below its initial value. Gas at the orbit
falls to $\BuoyLocalSigma\ {\rm g\,cm^{-2}}$, or $\BuoyLocalPercent\%$ of the
initial column at that same radius; the ordinary-only control retains 24.6\%.
The exterior trough retains $\BuoyOuterPercent\%$ at its minimum. Thus local
deposition clears the gas remaining close to Ganymede while preserving the
effective stall (Figure~\ref{fig:buoyancyevolution}).

At the first saved nonnegative total torque (94.1 yr, $a=17.6R_J$),
$\Gamma_L=-4.26\times10^{31}$ and $\Gamma_B=+4.39\times10^{31}\ {\rm dyn\,cm}$
sum to $+1.37\times10^{30}\ {\rm dyn\,cm}$ (Table~\ref{tab:buoyancy_torque}).
The outward buoyancy reaction tips the balance while ordinary torque remains
inward; its initial 1.98\% inward contribution does not characterize this state.
Over 500--1,000 yr, the mean buoyancy reaction remains outward, partly offsetting
ordinary inward torque. Local deposition also changes ordinary excitation
through the density profile; these torque contributions come from the same
coupled evolution, not from independent orbital calculations.

\begin{table}[htbp]\centering\small
\caption{Signed orbital torque in the buoyancy-added run, in $10^{31}\ {\rm dyn\,cm}$. The first reversal is the first saved nonnegative total torque following a negative record, not the orbital minimum. The late row is a time average over 500--1,000 yr; no single radius is assigned to it. The total is calculated before rounding the contributions.}
\label{tab:buoyancy_torque}
\begin{tabular}{lrrrrr}\toprule
State & $t$ [yr] & $a/R_J$ & $\Gamma_L$ & $\Gamma_B$ & $\Gamma_{\rm total}$\\\midrule
Initial & 0 & 20.0 & -77.8 & -1.54 & -79.3\\
First reversal & 94.1 & 17.6 & -4.26 & +4.39 & +0.137\\
Late mean & 500--1,000 & --- & -1.25 & +0.883 & -0.372\\
Final & $1.00\times10^{3}$ & 17.9 & -0.979 & +0.849 & -0.130\\
\bottomrule\end{tabular}
\end{table}

Over 500--1,000 yr, separate linear detrending gives residual rms excursions
$\BuoyRMS R_J$ with buoyancy versus $\ControlRMS R_J$ without, a reduction
by a factor of $\BuoyRMSFactor$; peak-to-peak ranges are $\BuoyPtp R_J$ and
$\ControlPtp R_J$. Both use a common 0.1-yr cadence. These are resolved secular
excursions, not damping eigenvalues. Together with
Figure~\ref{fig:buoyancyevolution}'s density history, they support repeated
inner--outer torque reversals as ordinary-only Ganymede moves relative to residual
near-orbit gas and adjacent depleted regions. Local deposition clears that gas and
smooths confinement.

In $[a-H_z,a+H_z]$, mass falls from $4.48\times10^{26}$ to $1.37\times10^{25}$ g
over 0--1,000 yr (Table~\ref{tab:buoyancy_mass}). Direct buoyancy transport
removes $4.75\times10^{26}$ g; viscosity returns $1.23\times10^{26}$ g, with
further net losses from Rayleigh transport and annulus motion. During 500--1,000
yr, viscosity and Rayleigh transport instead replenish \BuoyLateReturnPercent\%
of buoyancy removal, yet mass still declines. Rayleigh transport therefore
opposes late clearing, although it removes gas when integrated over the full
history. Ordinary deposition contributes zero through these saved annular
boundaries but still affects the global profile and orbit.

The final mass in this $2H_z$-wide annulus is \BuoyAnnulusPercent\% of the
initial-profile mass over the same final radial limits, compared with
\BuoyLocalPercent\% for the column at the orbit. Both show strong depletion. The
budget along the evolving orbit identifies buoyancy as the main direct source of
local gas removal, while including the coupled ordinary and Rayleigh response;
no source fraction is fitted. Appendix~\ref{app:annulus_reference} specifies the
reference mass and diagnostic closure.

\begin{table}[htbp]\centering\small
\caption{Mass changes in the moving annulus $[a-H_z,a+H_z]$ of the buoyancy-added run, in $10^{26}$ g. Negative values remove annular mass. Flux diagnostics are integrated over saved times; the residual is measured change minus their sum. Sums and residuals are calculated before rounding.}
\label{tab:buoyancy_mass}
\begin{tabular}{lrr}\toprule
Contribution & 0--1,000 yr & 500--1,000 yr\\\midrule
Ordinary deposition & 0 & 0\\
Local buoyancy & -4.75 & -0.951\\
Background viscosity & +1.23 & +0.483\\
Rayleigh transport & -0.556 & +0.255\\
Moving annulus edges & -0.295 & +0.000393\\
\midrule
Sum of contributions & -4.37 & -0.213\\
Measured annular change & -4.34 & -0.213\\
Diagnostic closure residual & +0.0267 & $+7.72\times10^{-6}$\\
\bottomrule\end{tabular}
\end{table}

The maximum eccentricity $\BuoyMaxE$ remains below $e=h$. Mass and added-transport
residuals are $\BuoyMassResidual$ and $\BuoyJResidual$ of initial disk
inventories; the full endpoint residual is $\BuoyFullJPercent\%$ of absolute
orbital angular-momentum change, including ordinary discretization and integration
errors. This mechanical result assumes unsaturated local deposition and low-$m$
residue extrapolation. Opposite redistribution inside the orbit-straddling grid
interval remains unresolved (Appendix~\ref{app:buoyancy_discrete});
Section~\ref{sec:introduction} gives the follow-up scope.

\section{Disk clearing and resonant migration with gap-edge redistribution}
\label{sec:pressure_edges}

Satellite torques and angular-momentum deposition displace gas, producing
Rossby-wave unstable pressure bumps at the edges of the cleared region. 
These bumps can form while
the edges remain Rayleigh stable, because Rayleigh stability does not require
a monotonic pressure profile. The existing Rayleigh adjustment therefore does
not suppress them. This motivates a separate pair of calculations that proactively
inhibits 
the formation of gap-edge pressure bumps through redistribution that 
conserves mass and angular momentum, for isolated Ganymede and for the two Callistos.
Here \emph{stalling} describes strong suppression of inward migration, whereas
an orbit that has \emph{stalled} has reached sustained radial confinement.

\subsection{Pressure-bump criterion and redistribution conserving mass and angular momentum}
\label{sec:pressure_criterion}

We adopt a simplified, deliberately permissive pressure trigger, allowing
redistribution without requiring demonstrated Rossby-wave instability. In the
inviscid, two-dimensional adiabatic theory of
\citet{Lovelace1999} and \citet{Li2000}, a radial extremum of the entropy-weighted inverse
vortensity is a necessary condition for instability. In our notation this
function is
\begin{equation}
  \mathcal L(R)=\frac{\Sigma\Omega_{\rm pb}}{\kappa_{\rm pb}^2}
  S^{2/\gamma},\qquad S=\frac{\Pi}{\Sigma^\gamma},
  \label{eq:rossby_function}
\end{equation}
where $\Pi$ is vertically integrated pressure, $S$ is an entropy proxy,
$\Omega_{\rm pb}$ is the pressure-balanced angular frequency, and
$\kappa_{\rm pb}$ is its radial epicyclic frequency. For the axisymmetric
background velocity $\boldsymbol u$ in the inertial frame, with azimuthal component
$u_\phi=R\Omega_{\rm pb}$, the vertical vorticity $\omega_z$ satisfies
\begin{align*}
  \omega_z\equiv(\nabla\times\boldsymbol u)_z
  &=\frac{1}{R}\frac{d(Ru_\phi)}{dR}
   =2\Omega_{\rm pb}+R\frac{d\Omega_{\rm pb}}{dR},\\
  \kappa_{\rm pb}^2
  &=\frac{1}{R^3}\frac{d(R^4\Omega_{\rm pb}^2)}{dR}
   =2\Omega_{\rm pb}\omega_z.
\end{align*}
Thus equation~\eqref{eq:rossby_function} is equivalently
$\mathcal L=\Sigma S^{2/\gamma}/(2\omega_z)$, one-half the entropy-weighted
inverse vortensity. Here $\omega_z$ includes the background rotation; it is not
the vorticity perturbation of an individual vortex.

Growth also depends on the feature's amplitude and width. Our heuristic targets
pressure maxima directly;
it does not solve the Rossby eigenvalue problem or enforce monotonic
$\mathcal L$.

Temperature remains prescribed, so
$\Pi=\Sigma c_{\rm iso}^2\propto\Sigma/R$ changes through redistribution of gas.
Here $c_{\rm iso}=c_{\rm ad}/\sqrt{\gamma}$ and
$H_{\rm iso}=c_{\rm iso}/\Omega_K$ are the isothermal sound speed and scale
height. The pressure slope is therefore
\begin{equation}
  \frac{d\ln\Pi}{d\ln R}=\frac{d\ln\Sigma}{d\ln R}-1.
  \label{eq:pressure_bump_slope}
\end{equation}
At a smooth pressure maximum the surface-density slope crosses 1 from above
to below; a density enhancement alone need not create a pressure maximum.
No evolving temperature maximum is introduced. On the grid, the trigger is a
rise followed by a fall in $\ln\Pi$, including a flat-topped maximum.

When a maximum develops at radius $R_p$, gas is redistributed within
$R_p\pm2H_{\rm iso}(R_p)$ to make pressure monotonic, conserving the window's
mass and Keplerian angular momentum while retaining positive density and
Rayleigh stability. The selected profile minimizes the mass-weighted squared
logarithmic density change, with nonnegative integrated stress
(Appendix~\ref{app:pressure_edges}). Overlapping windows are merged, and the
connected feature beside the inner drain is excluded. A merged window can
span both satellites and their shared gap: the final pair adjustment covers
$8.12$--$31.8R_J$. The ordinary excitation and deposition,
background viscosity and Rayleigh operator remain those of each respective
reference case. Both adjustments remain active in the pressure-adjusted runs.
We check the saved profiles for residual maxima after the adjustment.
Appendix~\ref{app:pressure_edges} specifies the tolerances, window
selection, fluxes that conserve mass and angular momentum, and the narrower
$H_{\rm iso}$ trial.
This calculation tests redistribution associated with gap-edge structure;
stochastic forcing by vortices remains the separate question introduced in
Section~\ref{sec:introduction}.

\subsection{Isolated Ganymede: continued clearing without pressure bumps}
\label{sec:ganymede_pressure}

\begin{figure}[!htbp]
  \centering
  \includegraphics[width=\textwidth]{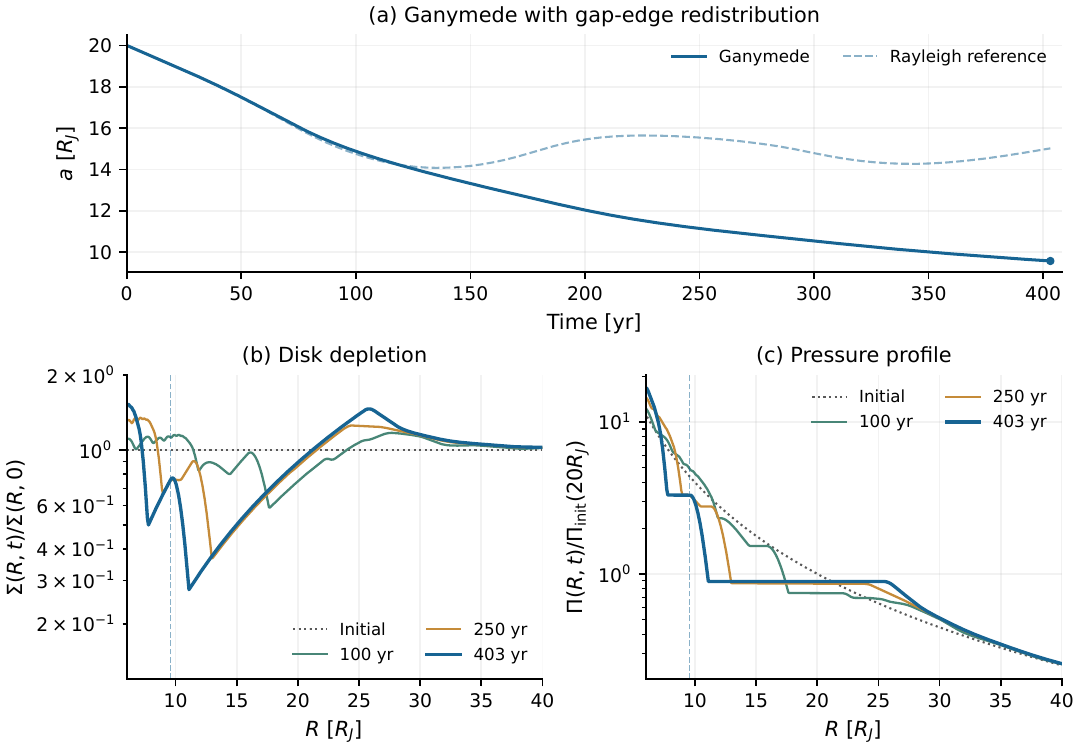}
  \caption{Isolated Ganymede with $2H_{\rm iso}$ gap-edge
  redistribution that conserves mass and angular momentum. (a) Semimajor axis, compared with the calibrated Rayleigh
  reference (dashed). (b) Surface density relative to its initial value at the
  same radius, at 100, 250 and 403 yr. (c) Vertically integrated pressure,
  divided by the fixed initial pressure at $20R_J$. The final exterior trough
  retains 27.5\% of the initial column, while pressure plateaus replace the
  gap-edge maxima. Vertical dashed lines in (b,c) mark the final orbit;
  dotted curves show the initial disk. The radial range excludes the inner
  drain. Profiles are saved states, without spatial smoothing.}
  \label{fig:pressure_ganymede}
\end{figure}

Figure~\ref{fig:pressure_ganymede} follows the calibrated Ganymede case with
this added redistribution and compares its orbit with the Rayleigh reference.
At the requested 403 yr endpoint, Ganymede has reached $9.57R_J$. Its mean
inward drift over the final 10 yr is
$7.31\times10^{-3}\,R_J\,{\rm yr}^{-1}$, compared with
$9.30\times10^{-3}\,R_J\,{\rm yr}^{-1}$ over the final 100 yr.
The exterior trough at $11.1R_J$ retains 27.5\% of its initial same-radius
column. The column is below half its initial value from $10.6$ to $15.0R_J$;
the contiguous region with at least 10\% depletion extends from $7.36$ to
$20.1R_J$. The column at the orbit retains 75.6\% of its initial local value,
showing why it must be distinguished from the exterior trough.
Across 640 saved profiles, 13 contain residual gap-pressure maxima, with
maximum pressure prominence only 0.026\%; none is detected at the endpoint
above the numerical tolerance.

Continued migration accompanied by clearing also prepares an
inherited depleted disk for later satellites
(Section~\ref{sec:early_transport}).

\subsection{Two Callistos: shared clearing and resonant capture}
\label{sec:pair_pressure}

\begin{figure}[!htbp]
  \centering
  \includegraphics[width=\textwidth]{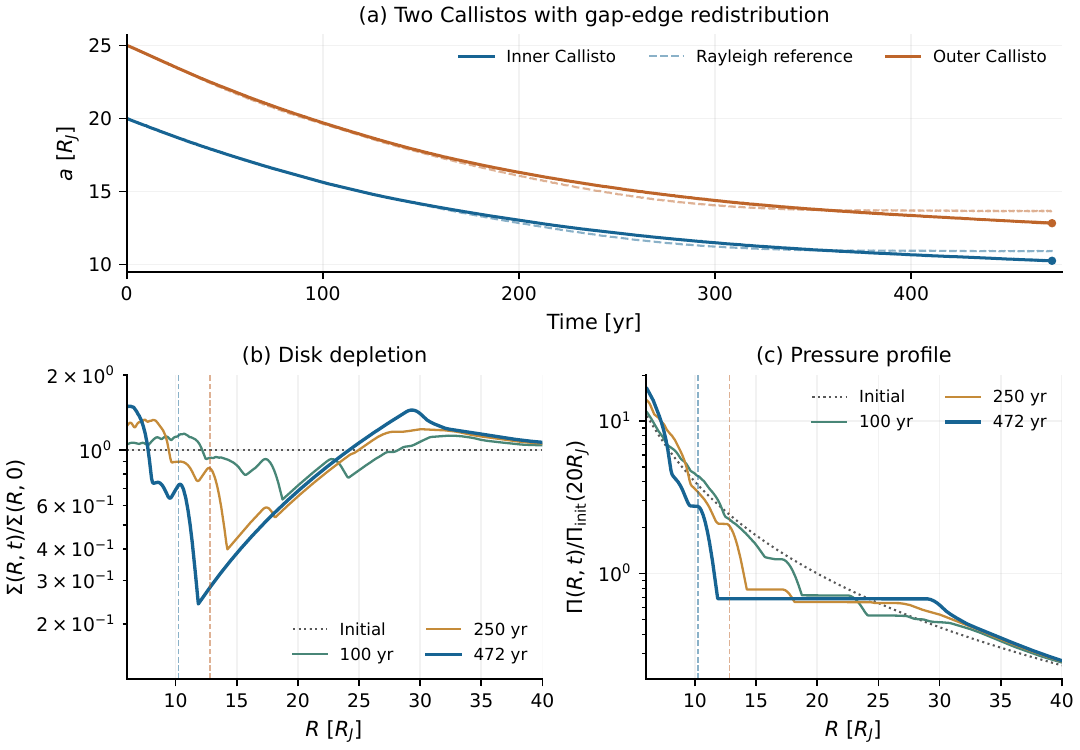}
  \caption{The two Callistos with $2H_{\rm iso}$ gap-edge
  redistribution that conserves mass and angular momentum. (a) Semimajor axes compared with the original Rayleigh
  pair (dashed curves). (b,c) Same-radius depletion fractions and pressure
  profiles at 100, 250 and 472 yr, using the conventions of
  Figure~\ref{fig:pressure_ganymede}. The satellites produce a shared trough
  while the adjustment suppresses gap-edge pressure maxima; the final
  intermoon minimum retains 24.1\% of its initial column. Vertical dashed
  lines mark the final satellite orbits. The average inward drift over
  400--470 yr is reduced by 89\% and 88\% from the respective initial rates.
  These interval averages do not describe an established limiting migration
  rate.}
  \label{fig:pressure_callistos}
\end{figure}

The two Callistos clear a shared depleted region and migrate together
in 7:5 resonance while retaining modest eccentricities. We repeat the reference
Callisto-pair calculation of Section~\ref{sec:pair_extension}, adding the
$2H_{\rm iso}$ pressure adjustment of Section~\ref{sec:pressure_edges}.
No additional finite-thickness multiplier or buoyancy source is introduced in
this matched pair comparison. Figure~\ref{fig:pressure_callistos} shows the
orbital evolution and the developing depleted region.
The initial period ratio $(25/20)^{3/2}$ is only $0.18\%$ below $7/5$, so
capture begins from a near-commensurate configuration.
Disk self-gravity and the contribution of the disk's axisymmetric gravitational
potential to apsidal precession are omitted.

For inner and outer mean longitudes $\lambda_{\rm in}$, $\lambda_{\rm out}$
and longitudes of periapse $\varpi_{\rm in}$, $\varpi_{\rm out}$, we show
one second-order 7:5 resonant angle and the apsidal separation:
\begin{align}
  \theta_2&=7\lambda_{\rm out}-5\lambda_{\rm in}
                  -\varpi_{\rm in}-\varpi_{\rm out},\nonumber\\*
  \Delta\varpi&=\varpi_{\rm out}-\varpi_{\rm in}.
  \label{eq:pair_pressure_display}
\end{align}

\begin{figure}[!htbp]
  \centering
  \includegraphics[width=\textwidth]{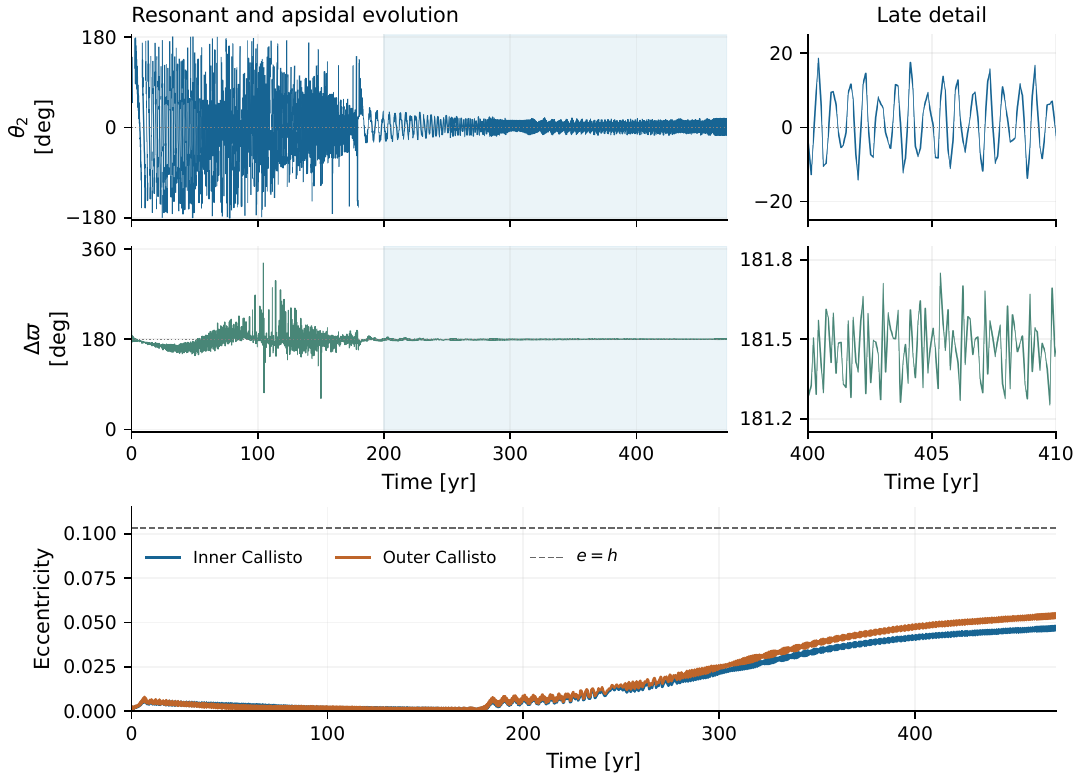}
  \caption{Resonant and apsidal locking of the pressure-adjusted Callisto pair.
  Top: the 7:5 angle $\theta_2$. Middle: the apsidal separation
  $\Delta\varpi$, with the dotted line at $180^\circ$ indicating
  anti-alignment. Both are defined in equation~\eqref{eq:pair_pressure_display}.
  The left panels show the full history, with the checked 200--472 yr
  libration interval shaded; the right panels enlarge 400--410 yr.
  The apsidal close-up resolves the small oscillations around $181.5^\circ$.
  The bottom panel shows both eccentricities remaining below 0.06 and the
  disk aspect ratio $h=0.103$. No temporal subsampling is applied; angle
  samples with undefined periapse are omitted as specified in
  Appendix~\ref{app:pressure_resonance}.
  The shaded interval does not assign an exact capture epoch; sampling and
  diagnostics for all three critical angles are given in
  Appendix~\ref{app:pressure_resonance}.}
  \label{fig:pressure_resonance}
\end{figure}

At the completed 472 yr endpoint, the semimajor axes are $10.3$ and
$12.8R_J$, and eccentricities are 0.0464 and 0.0533. The maximum eccentricities
throughout the saved history are 0.0481 and 0.0553, both below $h=0.103$.
Figure~\ref{fig:pressure_resonance} shows $\theta_2$, the apsidal separation,
and both eccentricities. The resonant angle librates over the checked
200--472 yr interval, while the periapses remain nearly anti-aligned.
Over 400--472 yr, the half peak-to-peak excursions are $17^\circ$ for
$\theta_2$ and $0.28^\circ$ for $\Delta\varpi$. Together these establish
resonant and apsidal locking. The three critical angles differ only by
$\Delta\varpi$, so their librations have nearly the same shape.
Appendix~\ref{app:pressure_resonance} retains all three definitions and
measurements.

Over the common 400--470 yr interval, fitted inward migration rates are
$5.84\times10^{-3}$ and $7.32\times10^{-3}\,R_J\,{\rm yr}^{-1}$, reductions
of 89\% and 88\% relative to the initial rates. The original pair, without
pressure adjustment, has rates $4.39\times10^{-4}$ and
$5.62\times10^{-4}\,R_J\,{\rm yr}^{-1}$ over the same interval, about 99\%
slowing. These are averages over the stated interval, not limiting migration
rates. The added redistribution allows more inward motion than the reference
treatment. The mean inward rates are
$5.92\times10^{-3}$ and $7.68\times10^{-3}\,R_J\,{\rm yr}^{-1}$ over the
final 100 yr and $5.38\times10^{-3}$ and
$6.45\times10^{-3}\,R_J\,{\rm yr}^{-1}$ over the final 10 yr.

The final intermoon trough has $\Sigma=8.12\times10^3\ {\rm g\,cm^{-2}}$ at
$11.9R_J$, 24.1\% of its initial same-radius column. The gas mass in the
annulus between the semimajor axes is 43.0\% of the initial mass in that same
annulus. The orbital columns retain 72.3\% and 28.1\%, respectively.
No resolved gap-edge pressure maximum remains at the endpoint; only one of the
1,562 saved profiles has a residual maximum under the final gap-selection
rule, with fractional pressure prominence $5.2\times10^{-9}$.
The drain feature is excluded throughout this diagnostic.

Together, the pressure-adjusted calculations show sustained disk clearing and
strong migration suppression while proactively avoiding gap-edge pressure
bumps. The Callisto pair also maintains a 7:5 resonance at modest
eccentricity. The adjustment permits more inward motion than the reference
treatment, and no orbital confinement is imposed. Both integrations ended
through normal saving because further evolution was computationally
expensive
(Appendix~\ref{app:pressure_edges}).

\FloatBarrier
\section{Conclusions}
\label{sec:conclusions}

We find that spatial excitation and launch-dependent shock onset permit tracking
extended gaps in circumplanetary disks. In agreement with Rafikov's stalling
criterion, we have shown that large satellites can stall in a dense,
low-viscosity disk. Radial confinement and gas clearing persist under different
treatments of the Rayleigh instability, including replacing an instantaneous
adjustment with finite transport beginning halfway to Rayleigh marginality, at
$\chi=0.5$ (Section~\ref{sec:gradient_sensitivity}). These results support
survival in the specified dense, low-viscosity circumplanetary disk.

Specifically, we found that an isolated Ganymede excavates the outer disk and
stalls near $15R_J$. Exterior gas depletion by itself supplies the required
torque deficit for Ganymede to stall, without enhanced inner feedback. This is a different stalling mechanism
from the one typically invoked for Earth-mass and super-Earth planets in protoplanetary disks, 
where the inner disk feedback is the main factor leading to a planetary stall.
Likewise, the Callisto satellite
pair clears a shared depleted region and stalls near $11$ and $14R_J$. However, the outer
satellite's eccentricity already exceeds the $e=h$ threshold during the stalling phase. Furthermore, a late satellite close encounter
disrupts the satellite orbits. Self-consistent eccentric disk--satellite coupling
and its effects on resonant evolution require follow-up modeling.

We treat angular-momentum-conserving inhibition of Rossby-wave unstable
regions at Rayleigh-stable gap edges and find that proactively 
avoiding pressure bumps
still allows extended clearing with strong migration suppression for
Ganymede and the two Callistos; however, as expected, our code dramatically 
slows down once a gap forms
and the satellites reconstitute the bumps essentially at each time step.
Over the final 10 yr of the 403-yr Ganymede run and the 472-yr two-Callisto run,
the mean inward migration speeds are 15\% and 10--11\% of their respective
initial values. At these endpoints, Ganymede's exterior trough and the pair's
intermoon trough retain 28\% and 24\% of their initial surface densities at
the same radii, respectively.
For the two Callistos, libration of all three critical angles demonstrates resonant
capture in a 7:5 resonance maintained over the 200--472 yr interval
(Appendix~\ref{app:pressure_resonance}). This run avoided the high-eccentricity
regime we encountered in our initial two-Callisto simulation, 
with satellite eccentricities
remaining below 0.06.

We also find that local angular momentum deposition
\citep{MosqueiraBuoyancy2026} clears the residual gas pile-up at the satellite's
orbit due to non-local shock deposition. 
Satellite radial oscillations shrink as a result. Unsaturated buoyancy
deposition is the main process removing gas from the annulus around the satellite's
orbit. Rayleigh redistribution also helps to erode this gas pile-up. The clearing of this 
residual gas explains the smoother satellite stalling behavior in the presence
of local deposition.

\begin{acknowledgments}
This paper was developed in collaboration with GPT-6 Pro and GPT-6 Astra. 
\end{acknowledgments}

\FloatBarrier
\appendix

\section{Finite-thickness excitation and its calibration}
\label{app:three_d}

The extension preserves resonance geometry, density sampling and pressure cutoff.
We distinguish softening length $b_{\rm soft}$, Laplace coefficient $\mathcal
B_m$, original coefficient $C_{m,\sigma}$ and correction $\mathcal F_{m,\sigma}$.

\subsection{Unsoftened modal forcing}
\label{app:modal_coefficient}

In this appendix only, $z=R/a$ is a dimensionless radius; physical height is
used for $z$ in Appendix~\ref{app:buoyancy_coupling}. Primes here denote
$d/dz$, and $\delta_{m1}$ is the Kronecker delta (one for $m=1$, zero otherwise).
The dimensionless potential coefficient $\phi_m$ and forcing coefficient
$C_{m,\sigma}$, evaluated at $z=R_{m,\sigma}/a$, are
\begin{align}
  \phi_m(z)&=-b_{1/2}^{(m)}(z)+\delta_{m1}z,\nonumber\\
  C_{m,\sigma}&=\frac{\pi^2 z^{3/2}[z\phi_m'(z)-2\sigma m k_m\phi_m(z)]^2}
  {3k_m[1+4(mh)^2]}.
  \label{eq:coefficient}
\end{align}
With $\theta$ the relative azimuth, the Laplace coefficient is
$b_{1/2}^{(m)}(z)=\pi^{-1}\int_0^{2\pi}\cos(m\theta)(1-2z\cos\theta+z^2)^{-1/2}d\theta$.
A 1,024-point Gauss--Legendre quadrature on $[0,\pi]$ evaluates the symmetric
integral and derivative. 
\subsection{Full angular kernel and fixed-mode derivative}

The finite-thickness approximation of \citet{MenouGoodman2004}, equations 13--18,
softens the direct gravitational potential on a scale related to the vertical
density distribution. We use
\begin{equation}
  b_{\rm soft}(R)=\eta H_z(R),\qquad H_z=\frac{c_{\rm iso}}{\Omega_K},\qquad
  c_{\rm iso}^2=\frac{k_BT}{2.3m_H}.
  \label{eq:3dheight}
\end{equation}
The vertical scale is distinguished from the adiabatic acoustic scale used in
$k_m$ and the pressure cutoff. For equation~\eqref{eq:disk}, $h_{\rm
iso}=H_z/R=h/\sqrt\gamma=0.0871$ is constant. More generally, a prescribed power
law $T\propto R^{-q_T}$ gives $H_z\propto R^{(3-q_T)/2}$ within the retained
Keplerian geometry.

Put $z=R/a$, $s(z)=b_{\rm soft}(az)/a$, and
$\mathcal D=1+z^2-2z\cos\theta+s(z)^2$. The coefficient and its radial derivative
are evaluated from
\begin{align}
  \mathcal B_m(z;\eta)&=\frac{2}{\pi}\int_0^\pi
  \frac{\cos(m\theta)}{\sqrt{\mathcal D}}\,d\theta,\label{eq:3dkernel}\\
  \mathcal B'_m(z;\eta)&=\frac{2}{\pi}\int_0^\pi
  \frac{\cos(m\theta)[\cos\theta-z-s(z)s'(z)]}
  {\mathcal D^{3/2}}\,d\theta.\label{eq:3dderivative}
\end{align}
The derivative holds $m$, $a$ and $\eta$ fixed but includes the prescribed radial
variation of $H_z$. Here $s=\eta h_{\rm iso}z$ and $ss'=\eta^2h_{\rm iso}^2z$. We
evaluate the full angular integral for every integer mode, including the lowest
modes, instead of the small-angle Bessel approximation. The unsoftened limit
$\eta=0$ recovers the original coefficient exactly within quadrature precision.

At each unchanged resonance $z_{m,\sigma}=(1+\sigma k_m/m)^{2/3}$, define
\begin{align}
  \phi_m^{\eta}&=-\mathcal B_m+\delta_{m1}z,\nonumber\\
  \Psi_{m,\sigma}^{\eta}&=z(\phi_m^{\eta})'-2\sigma m
  k_m\phi_m^{\eta},\label{eq:3dforcing}\\
  \mathcal
  F_{m,\sigma}(\eta)&=\frac{|\Psi_{m,\sigma}^{\eta}|^2}{|\Psi_{m,\sigma}^{0}|^2},
  &A_{m,\sigma}^{\eta}&=\mathcal F_{m,\sigma}(\eta)A_{m,\sigma}^{0}.
  \label{eq:3dfactor}
\end{align}
The indirect term and derivative are unchanged, pressure factors apply once and
resonance radii are uncorrected. Numerically, amplitudes use $|\Psi^\eta|^2$
directly with the retained prefactor to avoid division by tiny high-mode
amplitudes; ratios are saved separately. Both sides use one $\eta$, with
differences set by resonance geometry and forcing.

\subsection{One smooth-disk constraint on one parameter}

For the reference profile, $\Sigma(R_{m,\sigma})/\Sigma(a)=z_{m,\sigma}^{-p}$ with
$p=1$. The normalization and target are
\begin{align}
  \Gamma_{0,\rm
  ad}&=\Sigma(a)a^4\Omega_K^2(a)(\mu/h)^2,\label{eq:3dnormalization}\\
  \frac{\Gamma_{\rm spec}(\eta_\star)}{\Gamma_{0,\rm ad}}
  &=\frac{\sum_m[A_{m,-}^{\eta_\star}-A_{m,+}^{\eta_\star}]}{\Gamma_{0,\rm ad}}
  =-(2.34-0.1p+1.5q_T)=-3.74.
  \label{eq:3dtarget}
\end{align}
This is the Lindblad expression of \citet{JimenezMasset2017}, equations 39--40, in
the slow-diffusion limit; ordinary corotation is excluded. Their factor $1/\gamma$
is already included through $h=h_{\rm ad}=\sqrt\gamma h_{\rm iso}$ in
equation~\eqref{eq:3dnormalization}. Applying it again would double count the
conversion. In these units an individual reference amplitude is
$C_{m,\sigma}z_{m,\sigma}^{-p}h^2$. All reference resonances at $a=20R_J$ lie
inside the modeled disk, so no boundary extrapolation enters this calibration.

\begin{figure}[!htbp]
  \centering
  \includegraphics[width=\textwidth]{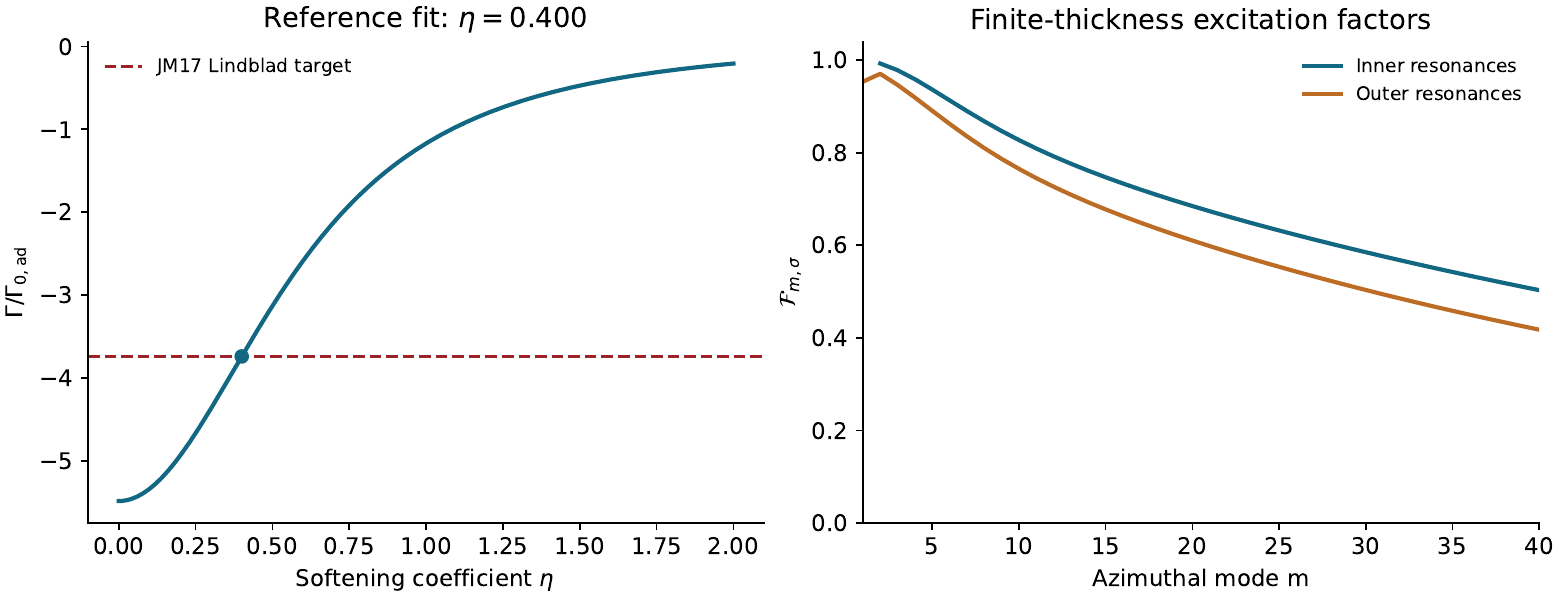}
  \caption{Static calibration on the undepleted reference disk. Left: the signed
  satellite torque as the common softening coefficient varies, with the
  Jim\'enez \& Masset target and fitted $\eta_\star$ marked. Right: the resulting
  factors at individual inner and outer resonances. The same parameter is used for
  both sides. The original resonance positions and pressure factors are retained;
  no orbital evolution is performed.}
  \label{fig:3dcalibration}
\end{figure}

The fitted value is
\begin{equation}
  \eta_\star=0.400,\qquad
  b_{\rm soft}/R=0.0349.
  \label{eq:3deta}
\end{equation}
Appendix~\ref{app:reproduction} retains the full-precision fitted coefficient for
numerical reproduction; it does not imply corresponding physical accuracy.
Table~\ref{tab:3dsums} reports the one-sided sums and net difference;
Table~\ref{tab:3dfactors} gives representative factors.
Figure~\ref{fig:3dcalibration} displays the root and spectral changes.

\begin{table}[htbp]
\centering
\caption{Static smooth-disk launch sums and satellite torque in units of $\Gamma_{0,\rm ad}$. Positive one-sided amplitudes are listed separately. The net is calculated before rounding.}
\label{tab:3dsums}
\begin{tabular}{lrr}\toprule
Quantity & Original & Calibrated kernel \\ \midrule
Inner launch sum & 8.95 & 7.29 \\
Outer launch sum & 14.4 & 11.0 \\
Total excitation & 23.4 & 18.3 \\
Satellite torque & -5.48 & -3.74 \\
\bottomrule\end{tabular}
\end{table}

\begin{table}[htbp]
\centering
\caption{Representative multiplicative factors $\mathcal F_{m,\sigma}$ from the same fitted $\eta_\star$. The outer $m=1$ forcing retains its indirect term.}
\label{tab:3dfactors}
\begin{tabular}{rrr}\toprule
Mode $m$ & Inner factor & Outer factor \\ \midrule
1 & --- & 0.954 \\
2 & 0.993 & 0.971 \\
5 & 0.937 & 0.891 \\
10 & 0.827 & 0.765 \\
20 & 0.685 & 0.610 \\
40 & 0.503 & 0.418 \\
\bottomrule\end{tabular}
\end{table}

Writing $S=A_-+A_+$ and $\epsilon=(A_+-A_-)/S$ as in
equation~\eqref{eq:decomposition}, the static comparison gives
\begin{equation}
  \frac{S^{\eta}}{S^0}=0.784,\qquad
  \frac{\epsilon^{\eta}}{\epsilon^0}=0.870,\qquad
  \frac{\Gamma^{\eta}}{\Gamma^0}=0.682.
  \label{eq:3dseparation}
\end{equation}
The net reduction combines weaker excitation (21.6\%) and stronger cancellation
(net magnitude reduced 31.8\%). Inner/outer $m=2$ retain 99.3\%/97.1\%; higher
modes are attenuated more. Ninety-nine percent of corrected excitation lies in
$m\leq34$, versus $m\leq39$ originally. These are calibration outputs.

\subsection{Numerical evaluation and interpretation}

The calculation retains outer $m=1,\ldots,256$ and inner $m=2,\ldots,256$, and
uses 1,024 Gauss--Legendre nodes on $[0,\pi]$. The full-precision constants and
resulting aspect ratio are specified in Appendix~\ref{app:reproduction}. A scan of
$0\leq\eta\leq2$ in increments of 0.01 finds one crossing, between 0.40 and 0.41;
Brent's method refines it with absolute parameter tolerance $10^{-13}$. The
sampled net torque is strictly increasing on that interval. This establishes the
detected root within the stated search, not a theorem of global uniqueness.

The zero-softening amplitudes reproduce the frozen spectrum to $5.2\times10^{-16}$
of its peak amplitude. Repeating the angular integration with 2,048 nodes changes
the net torque by $3.4\times10^{-12}$ fractionally and either one-sided sum by
less than $5.5\times10^{-12}$. A five-point finite-difference derivative with step
$\Delta z=10^{-5}$, allowing $s(z)$ to vary, agrees with
equation~\eqref{eq:3dderivative} to $1.5\times10^{-11}$ fractionally at modes 1,
2, 5, 10, 20 and 40 on the available sides. These checks establish numerical
accuracy of the specified static calculation, not physical accuracy of its
three-dimensional approximation.

At fixed $\eta_\star$, $h$, temperature slope, resonance geometry and
$\Sigma(a)$, vary the smooth profile as $\Sigma(R)=\Sigma(a)(R/a)^{-p}$. The
same frozen modal coefficients give
\begin{equation}
  \left.\frac{\partial(\Gamma_{\rm spec}/\Gamma_{0,\rm ad})}{\partial
  p}\right|_{p=1}
  =\sum_{m,\sigma}\sigma\ln z_{m,\sigma}\,
  \widehat A_{m,\sigma}(1)=2.21,
  \label{eq:slope_diagnostic}
\end{equation}
where $\sigma=-1,+1$ denotes inner/outer branches and $\widehat
A=A/\Gamma_{0,\rm ad}$. The target derivative is 0.1; a centered $\Delta
p=10^{-4}$ difference independently reproduces the modal value. Thus the kernel
determines the response to density gradients; matching the net torque alone does
not fix this response. It can affect confinement radii and radial excursions.
Calibration stays fixed without side factors or refitting.

One scalar fit does not determine a three-dimensional torque for each mode or
validate one-sided fluxes in gaps. \citet{Masset2011}, Section 8.2, noted an
unresolved discrepancy with \citet{MenouGoodman2004}'s differential coefficients.
We adopt the latter's softened kernel, not their net-torque fit, calibrating to
the later target. Full angular integration removes only the small-angle
approximation.

\subsection{Use in an evolution and the separate shock-amplitude choice}

During evolution $\eta_\star$ stays fixed, softening follows prescribed height,
and corrected amplitudes sample current launch densities. Their inner-minus-outer
sum sets the ordinary satellite torque; individual signed contributions enter
deposition and escape. No local target or side correction is reimposed in a gap.
Modes $m\leq2$ still escape.

The amplitude-to-shock mapping is a distinct modeling choice. The present shock
criterion, equation~\eqref{eq:shock}, contains $\mu^2$ rather than the
independently integrated launch flux. At fixed disturbance shape, multiplying that
flux by $\mathcal F_{m,\sigma}$ multiplies its amplitude by $\sqrt{\mathcal
F_{m,\sigma}}$. The completed 3D comparison applies this change in the existing
shock estimate:
\begin{equation}
  \mu^2\longrightarrow\mathcal F_{m,\sigma}\mu^2,
  \qquad\mu_{\rm eff,m,\sigma}=\sqrt{\mathcal F_{m,\sigma}}\,\mu
  \quad\hbox{in the shock-amplitude input only}.
  \label{eq:3dshockchoice}
\end{equation}
The mode/side factor enters immediate- and delayed-shock tests only. Physical
mass, gravity, resonances and both baseline histories are unchanged; changes to a
superposed wake's shape are not calculated. The prescribed tail and shock law are
retained, with this choice fixed in the buoyancy comparison. Background viscosity,
Rayleigh transport and buoyancy are separate.

\subsection{Low-mode contributions in the saved evolving profiles}

The $m\leq2$ ordinary contributions retain their torque but escape without
deposition. In the smooth reference spectrum they carry 3.23\% of total excitation
and 9.24\% of the initial net inward torque. Table~\ref{tab:low_modes} evaluates
the same contributions on the saved profiles at each first reversal and in late
confinement. Every mode samples the actual evolving density at its own resonance,
using the same interpolation as the evolution. The independently reconstructed
one-sided sums reproduce the saved totals within $3\times10^{-13}$ relatively.

\begin{table}[htbp]\centering\small
\caption{Ordinary $m\leq2$ launched angular-momentum fluxes in $10^{31}\ {\rm dyn\,cm}$ and their fractions of the corresponding full one-sided flux. First reversal denotes the first saved nonnegative total orbital torque: 136 yr without buoyancy and 94.1 yr with it (Table~\ref{tab:buoyancy_torque}). The late rows use ratios of time-integrated fluxes over 500--1,000 yr; they are not normalized by the nearly vanishing net torque.}
\label{tab:low_modes}
\begin{tabular}{llrrrr}\toprule
Case & State & $A_{-,\leq2}$ & $A_{+,\leq2}$ & Inner [\%] & Outer [\%]\\\midrule
Ordinary only & First reversal & 3.07 & 7.93 & 2.30 & 5.96\\
Ordinary only & Late mean & 3.25 & 7.58 & 7.14 & 16.6\\
Ordinary only & 1,000 yr & 3.02 & 8.02 & 7.31 & 22.5\\
With buoyancy & First reversal & 2.90 & 9.78 & 1.93 & 6.32\\
With buoyancy & Late mean & 2.82 & 7.42 & 5.98 & 15.3\\
With buoyancy & 1,000 yr & 2.67 & 6.38 & 6.93 & 16.1\\
\bottomrule\end{tabular}
\end{table}

During 500--1,000 yr the low modes carry 7.14\% of the mean inner and 16.6\% of
the mean outer ordinary flux in the buoyancy-free case; with buoyancy these
fractions are 5.98\% and 15.3\%. At the ordinary-only first reversal,
$\Gamma_{\leq2}=\LowFirstNet\ {\rm dyn\,cm}$ remains inward, while
$\Gamma_{m>2}=+\HighFirstNet\ {\rm dyn\,cm}$ overcomes it to give the small
positive total. The mean low-mode torque over 500--1,000 yr is still inward,
$\LowLateNet\ {\rm dyn\,cm}$. Thus the escaping modes continue to pull inward;
the higher-mode balance supplies the ordinary-only reversal. This diagnostic
does not establish that the result is independent of the escape cutoff: a small
net torque can depend on changes much smaller than either full one-sided sum. No
normalization by that almost-zero net torque is used.

\section{Specified local buoyancy coupling}
\label{app:buoyancy_coupling}

The matched experiment (Section~\ref{sec:buoyancy_summary}) adds excitation and
local deposition without reassigning ordinary flux. All ordinary excitation,
propagation, transport and boundary prescriptions remain fixed. Only the direct
buoyancy exchange is included: the coorbital vortensity evolution and associated
dynamical corotation torque discussed in Section~\ref{sec:buoyancy_summary} are
outside this coupling.

\subsection{Analytical source and normalization}

Here $z$ is physical height above the midplane. For satellite $i$, let
$\Omega_i=\Omega_K(a_i)$ and write the equilibrium volume density as
$\rho_0(R,z,t)=\Sigma(R,t)w_z(R,z)$, with normalized vertical shape
$\int w_z\,dz=1$. This weight differs from the deposition-tail width $w$.
The vertical shape and temperature are prescribed. Let $g_z$ denote the magnitude
of vertical gravity and $N_z$ the vertical buoyancy frequency. For the
vertically isothermal structure used to define
$H_z=c_{\rm iso}/\Omega_K$, the corresponding adiabatic stratification is
\begin{equation}
  w_z=\frac{e^{-z^2/(2H_z^2)}}{\sqrt{2\pi}H_z},\qquad
  g_z=\Omega_K^2|z|,\qquad
  N_z^2=\frac{\gamma-1}{\gamma}\Omega_K^2\frac{z^2}{H_z^2}.
  \label{eq:bvertical}
\end{equation}
An isothermal vertical equilibrium therefore does not remove buoyancy when
disturbances are adiabatic. The initial $\Sigma$ and $T$ are those of
equation~\eqref{eq:disk}, not the companion paper's different reference column.
Prescribing this structure isolates mechanical redistribution; it does not evolve
the retained heat.

We use the candidate asymptotic-residue source developed by
\citet{MosqueiraBuoyancy2026} from \citet{LubowZhu2014}. To fix its Fourier
normalization, let $\phi$ and $\phi_i$ be the disk and satellite azimuths,
$\vartheta=\phi-\phi_i$ their difference, and $\Phi_{s,i}$ the satellite's direct
potential per unit mass. Its axisymmetric term is $\Phi_{0i}$ and its Fourier
coefficients are $\Phi_{mi}$:
\begin{align}
  \Phi_{s,i}&=-\frac{GM_i}{\sqrt{R^2+a_i^2-2Ra_i\cos\vartheta+z^2}},\nonumber\\
  \Phi_{s,i}&=\Phi_{0i}+\sum_{m\geq1}\Re[\Phi_{mi}e^{im\vartheta}],&
  \Phi_{mi}&=\frac{1}{\pi}\int_0^{2\pi}\Phi_{s,i}\cos(m\vartheta)\,d\vartheta.
  \label{eq:bpotential}
\end{align}
This candidate retains the companion's direct-potential forcing for buoyancy. The
height dependence is evaluated explicitly; neither the acoustic softening length
nor its modal attenuation factor is applied again. The ordinary excitation,
including its existing indirect term, is unchanged.

Define $\omega_{mi}=m[\Omega(R,z)-\Omega_i]$ and $F_{mi}=N_z^2-\omega_{mi}^2$. The
signed excitation torque per unit radius on the gas, with $\delta$ denoting the
Dirac delta function, is
\begin{equation}
  E_{B,i}(R)=-\pi^2\sum_{m\geq1}mR\int dz\,
  \rho_0|\Phi_{mi}|^2\frac{N_z^4}{g_z^2}
  \operatorname{sgn}(\omega_{mi})\,\delta(F_{mi}).
  \label{eq:bsource}
\end{equation}
The intended reduction retains $\Omega=\Omega_K(R)$, consistent with the
prescribed rotation in the ordinary excitation. The general form above also states
where a different prescribed rotation would enter. With
$\widehat\Phi_{mi}=\Phi_{mi}/\mu_i$,
equations~\eqref{eq:bvertical}--\eqref{eq:bsource} define an absolute signed
kernel through
\begin{equation}
  D_{B,i}=E_{B,i}=\mu_i^2\Sigma(R,t)\mathcal K_{B,i}(R;a_i),
  \label{eq:bkernel}
\end{equation}
where $\mathcal K_{B,i}$ is equation~\eqref{eq:bsource} with $\rho_0$ replaced by
$w_z$ and $\Phi_{mi}$ by $\widehat\Phi_{mi}$. For the adopted constant aspect ratio
it is tabulated in the dimensionless radius $R/a_i$ and rescaled as the satellite
moves. The evolving density is sampled where the buoyancy resonance acts. There is
no fitted torque fraction or normalization to a desired gap depth.

The equality $D_B=E_B$ assumes sustained unsaturated local deposition and
negligible disturbance storage. Localized excitation motivates local deposition
but does not establish indefinite unsaturation; extrapolating the
large-wavenumber pressure residue to low $m$ is heuristic. Ordinary $m\leq2$
escape concerns a separate acoustic approximation. The implemented kernel
uses the spectrum and quadratures below.

\subsection{Resonance surfaces and signed cell integrals}

At simple radial roots $R_q(z)$ of $F_{mi}=0$, with $q$ indexing roots and
derivatives taken at fixed physical height,
\begin{equation}
  \delta(F_{mi})=\sum_q
  \frac{\delta[R-R_q(z)]}
  {\left|\partial_R N_z^2-2\omega_{mi}m\partial_R\Omega\right|_{R_q,z}}.
  \label{eq:bjacobian}
\end{equation}
This Jacobian fixes the delta-function normalization. A quadrature using
$z/H_z(R)$ must still evaluate the derivative at fixed $z$. Degenerate roots
require their limiting treatment rather than division by a vanishing derivative.
In the local Keplerian limit the radial offsets satisfy
\begin{equation}
  |R_{B,m}(z)-a_i|\simeq\frac{2a_i}{3m}\frac{N_z(z)}{\Omega_i}.
  \label{eq:boffset}
\end{equation}
Integrating over height and summing over harmonics therefore gives a radial
distribution, rather than a chosen one-cell damping width. For the usual inner
and outer branches the gas torques are negative and positive, respectively.

For radial cell $j$ bounded by faces $R_{j-1/2}$ and $R_{j+1/2}$, the module
supplies
\begin{equation}
  \mathcal T^B_{ij}=\mu_i^2\int_{R_{j-1/2}}^{R_{j+1/2}}
  \Sigma(R,t)\mathcal K_{B,i}(R;a_i)\,dR,\qquad
  \Gamma_{B,i}=-\sum_j\mathcal T^B_{ij}.
  \label{eq:bcell}
\end{equation}
Cell-integrated roots retain signed inner/outer exchanges separately; their sum
supplies orbital reaction. The mesh combines opposite contributions within an
orbit-straddling dual interval (Section~\ref{app:buoyancy_discrete}); separate
accounting does not resolve subgrid transport or demonstrate mesh convergence.

\subsection{Evaluated kernel and spatial quadrature}
\label{app:buoyancy_numerics}

For the adopted Gaussian structure and Keplerian rotation, the height roots are
explicit. Let $q=R/a_i$, $h_z=H_z/R=h/\sqrt\gamma$,
$\beta_z=(\gamma-1)/\gamma$ and $\zeta=|z|/a_i$. For resonant height $z_m$,
define $u_m=|z_m|/H_z$ and the dimensionless potential coefficient $b_m$:
\begin{equation}
  u_m=\frac{|z_m|}{H_z}=\frac{m|1-q^{3/2}|}{\sqrt{\beta_z}},\qquad
  b_m(q,\zeta)=\frac{1}{\pi}\int_0^{2\pi}
  \frac{\cos(m\vartheta)\,d\vartheta}{\sqrt{1+q^2-2q\cos\vartheta+\zeta^2}}.
\end{equation}
At a height root, $\zeta=h_zq u_m$. Integrating both height roots in
equation~\eqref{eq:bsource} gives
\begin{align}
  d\mathcal T_{B,m}
  &=\frac{\Sigma_{\rm ref}GM_i^2a_i}{M_J}\,
  s(q)K_m(q)\,dq,\qquad s=\Sigma/\Sigma_{\rm ref},\\
  K_m(q)&=\operatorname{sgn}(q-1)
  \frac{\pi^2\beta_z}{\sqrt{2\pi}h_z^2}
  m q^2 u_m e^{-u_m^2/2}
  b_m^2(q,h_zq u_m).
  \label{eq:bevaluated}
\end{align}
Here $a_i$ is in physical units and $\Sigma_{\rm ref}=2\times10^4\ {\rm
g\,cm^{-2}}$. Exact integration over the height roots, including their Jacobian,
eliminates radial root finding for this rotation; independent resonance-height
quadrature checks the one-sided sums. This is equation~\eqref{eq:bsource}'s
evaluated reduction, not a fitted formula.

We retain $m=1,\ldots,512$ and $|z|/H_z<8$. The direct point potential is
unsoftened. Its Fourier coefficients are evaluated through toroidal-function
recurrences: with $\chi_\Phi=1+[(q-1)^2+\zeta^2]/(2q)$, use forward recurrence for
$m\operatorname{arcosh}\chi_\Phi\leq2$ and the stable Miller recurrence otherwise.
This numerical switch changes the evaluation method, not the potential. Five
independent angular-integral checks spanning $m=1$--512 agree to better than
$1.5\times10^{-11}$ relatively. The local buoyancy source includes the low
harmonics under the stated residue extrapolation; the ordinary acoustic $m\leq2$
escape rule remains unchanged.

For each side, the summed kernel is stored at 4,097 logarithmically spaced offsets
$|q-1|$: from $10^{-10}$ to 0.98 inward and to 3 outward. These bounds cover the
nonzero, height-truncated source throughout the guarded orbit domain. Positive
magnitudes use shape-preserving cubic interpolation in logarithmic coordinates,
with the two signs retained separately. The finite-spectrum corotation limit
vanishes; the omitted integral below $10^{-10}$ is negligible and does not define
a physical softening radius. Radial integrals use 12-point Gauss--Legendre
quadrature with subdivisions $\Delta\ln|q-1|\leq0.3$, split at the satellite, and
the evolving piecewise-linear surface density. Thus excitation samples its actual
spatial distribution. The fixed gas grid still limits the resolved radial
structure.

Doubling the table sampling to 8,193 points changes the represented source weights
by $1.7\times10^{-10}$ in relative $L^1$ norm. Independent height and radial
integrals agree within $6\times10^{-6}$. Modes 257--512 contribute approximately
0.86\% of the initial one-sided buoyancy integrals. This records the finite
harmonic cutoff; it is not a validation of the low-$m$ residue approximation or an
assertion of infinite-spectrum convergence. No thermal feedback coefficient,
density floor, torque fraction or additional acoustic attenuation enters this
calculation.

\subsection{Conservative coupling and mechanical diagnostics}

With $D_L=\sum_iD_{L,i}$ and $D_B=\sum_iD_{B,i}$, the retained transport equation
becomes
\begin{align}
  F_M&=\frac{D_L+D_B-\partial_R(G_\nu+G_R)}{\ell_K'},&
  \Gamma_i&=\Gamma_{L,i}+\Gamma_{B,i},\label{eq:btransport}\\
  \left.\partial_t\Sigma\right|_B
  &=-\frac{1}{2\pi R}\partial_R\left(\frac{D_B}{\ell_K'}\right).
  \label{eq:bdirect}
\end{align}
The source uses the converged trial density and the same accepted signed
exchanges/time weights for gas flux and opposite orbital impulse, without direct
mass removal or a separately computed orbital torque. Boundary exchanges and
ordinary residuals remain in the full ledger. The unchanged Rayleigh constraint
responds to the new density; buoyancy is not restricted to unstable cells or
fitted to compensate it. Vertical $N_z^2$ and radial $\kappa_{\rm pb}^2$ are
distinct.

For the prescribed Keplerian rotation the vertically integrated heating rate
per unit disk area is
\begin{equation}
  q^+_{B,i}=\frac{\Omega_i-\Omega_K(R)}{2\pi R}D_{B,i}.
  \label{eq:bheat}
\end{equation}
A source retaining vertical shear instead requires height integration of pattern
work minus circular-orbit work before this reduction. The mechanical comparison
records this heating without changing the prescribed temperature. The companion's
retained-heating feedback coefficients are consequently not imported.

To measure clearing near the orbit, choose an annulus with edges $R_-(t)$ and
$R_+(t)$ and record its gas mass $M_{\rm ann}$. Its exact continuum budget is
\begin{equation}
  \frac{dM_{\rm ann}}{dt}
  =F_M(R_-)-F_M(R_+)
  +2\pi R_+\Sigma(R_+)\dot R_+
  -2\pi R_-\Sigma(R_-)\dot R_-.
  \label{eq:bannulus}
\end{equation}
At each edge, separate the four contributions $D_L/\ell_K'$, $D_B/\ell_K'$,
$-G_\nu'/\ell_K'$ and $-G_R'/\ell_K'$. This identifies direct buoyancy transport
and changed Rayleigh redistribution on a given state; differences between evolved
histories include their coupled feedback. The signed inner and outer torque
integrals, local gas inventory, ordinary torque and orbit should accompany this
budget. A small net $\Gamma_B$ need not imply weak clearing because
equation~\eqref{eq:bdirect} depends on the source's spatial variation.

\subsection{Discrete exchange, time integration and saved diagnostics}
\label{app:buoyancy_discrete}

For an internal gas face between nodes $j$ and $j+1$, the code integrates
equation~\eqref{eq:bevaluated} over the dual interval $[R_j,R_{j+1}]$, obtaining
$\mathcal T^B_{j+1/2}$, and uses
\begin{equation}
  F^B_{M,j+1/2}=\frac{\mathcal T^B_{j+1/2}}{\ell_{K,j+1}-\ell_{K,j}}.
  \label{eq:bdiscreteflux}
\end{equation}
The boundary intervals end at the physical gas face and adjacent evolved node,
using their actual difference in $\ell_K$. Inner and outer source integrals are
evaluated separately and saved before forming the signed face flux. Let $M_j$ be the gas mass in evolved cell $j$, $f$ index gas faces, and
$F_J^B$ denote the outward angular-momentum flux associated with buoyancy-driven
mass transport. The discrete mass update satisfies
\begin{equation}
  \sum_j\ell_{K,j}\left.\frac{dM_j}{dt}\right|_B
  +F^B_{J,\rm out}-F^B_{J,\rm in}
  =\sum_f\mathcal T^B_f.
\end{equation}
This is the compatible discrete realization of equation~\eqref{eq:bdirect}. The
finite mesh averages the source within each dual interval; it does not resolve
opposite motions on subgrid scales inside an interval crossing the orbit. Only
buoyancy excitation within the modeled gas domain is exchanged with the satellite
in this comparison.

The cumulative buoyancy gas angular momentum $J^B$ is advanced with the accepted
backward-Euler/BDF2 recurrence. Here $n$ indexes accepted states and
$\Delta t_n=t_{n+1}-t_n$. For BDF2 with $q_t=\Delta t_n/\Delta t_{n-1}$ and
$a_0=(1+2q_t)/(1+q_t)$,
\begin{equation}
  J^B_{n+1}=\frac{(1+q_t)J^B_n-q_t^2J^B_{n-1}/(1+q_t)
  +\Delta t_n\sum_f\mathcal T^B_{f,n+1}}{a_0}.
\end{equation}
The orbit receives $-(J^B_{n+1}-J^B_n)$ as its additional impulse. This replaces
only the added buoyancy term's time weighting; the original ordinary torque
treatment is retained. The existing orbit interface applies a planetocentric
specific torque. For the isolated body, multiplying the new barycentric reaction
by $(M_J+M_i)/M_J$ at that interface gives the prescribed change in the
barycentric orbital inventory. This coordinate conversion is not a physical
excitation correction. Its standalone finite-step check has relative impulse error
$1.8\times10^{-6}$ at the retained orbital resolution. The full coupled budget
records the remaining ordinary discretization and orbital errors.

The solve retains the grid of Section~\ref{sec:disk} and the tolerance,
iteration limit and maximum step of Appendix~\ref{app:integration}. Its source Jacobian uses the same piecewise-linear density. The
first implicit step has mass and angular-momentum residuals below
$8\times10^{-15}$ of initial inventories, distinct from full-run residuals. The
$e\geq h$, positivity and domain guards remain active without a density floor.
Checkpoints add $J^B$, its previous multistep value and signed face exchanges to
the Cartesian state.

The moving annulus is $[a-H_z(a),a+H_z(a)]$, clipped to the physical gas domain if
needed. Each saved state contains its mass, the four outward-positive transport
contributions at both edges, and the edge-motion terms in
equation~\eqref{eq:bannulus}. These diagnostic edge fluxes use linear
interpolation between represented faces; whole-domain budgets use the actual
discrete fluxes. Heating is integrated as $(\Omega_i-\Omega_K)d\mathcal T_B$ and
recorded without changing temperature. Thus the output separates direct local
deposition from the accompanying ordinary and Rayleigh response.

The ordinary calibration and shock-amplitude choice remain fixed.
\citet{JimenezMasset2017}'s temperature-gradient coefficient comes from locally
isothermal calculations; slow-diffusion sound-speed rescaling does not add this
adiabatic buoyancy source. No suppression fraction, excursion damping or final
orbit is imposed.

\subsection{Finite-annulus reference and diagnostic closure}
\label{app:annulus_reference}

We distinguish depletion at the orbit from clearing across a finite region. For
the initial $R^{-1}$ profile, the reference mass over the \emph{same final}
radial limits is
\begin{equation}
  M_{\rm ann,ref}(a)=\int_{a-H_z}^{a+H_z}2\pi R\Sigma_{\rm init}(R)\,dR
  =4\pi aH_z\Sigma_{\rm init}(a)=\BuoyAnnulusRefMass\ {\rm g},
  \label{eq:bannulus_reference}
\end{equation}
with $H_z/a=h/\sqrt\gamma$ and $a=17.9R_J$. The final annular mass is
\BuoyAnnulusPercent\% of this reference, compared with \BuoyLocalPercent\% for
the column at the orbit. Both measurements indicate strong depletion; the
annular ratio measures gas throughout the surrounding $2H_z$-wide interval.

This budget follows the evolving orbit and identifies buoyancy as the main
direct source of local gas removal, while including the coupled ordinary and
Rayleigh response. Trapezoidal integration of interpolated saved fluxes closes
annular loss to 0.615\% over the full history and 0.00362\% late; these differ
from the discrete global residual. No source fraction is fitted.

\section{Numerical specifications and analysis conventions}
\label{app:reproduction}

\subsection{Numerical setup and sampling}

The eight histories in Table~\ref{tab:case_map} use the common disk, masses,
boundaries and 801-node mesh specified in
Sections~\ref{sec:disk}--\ref{sec:numerics}. Appendices~\ref{app:three_d} and
\ref{app:buoyancy_coupling} give the calibration and source quadratures;
Appendix~\ref{app:rayleigh} gives the stress construction, integration scheme,
tolerances and guards. The finite-stress cases use
$(\chi_{\rm onset},\chi_{\rm cap})=(0.5,0)$ and $(1,0.5)$ with
$\alpha_{\rm extra,cap}=10^{-3}$. No density floor is used. Quadrature and
budget checks do not establish gas-mesh convergence; that assessment is reserved
for the forthcoming study outlined in Section~\ref{sec:introduction}.

Lengths are measured in $R_J$, time in 365.25-day years, and gas surface density
in units of $\Sigma_{\rm ref}=2\times10^4\ {\rm g\,cm^{-2}}$. The additional
Rayleigh stress at the 799 interior grid nodes uses the normalization
\begin{equation}
  G_{R,\rm unit}=2\pi\Sigma_{\rm ref}R_J^2\sqrt{GM_JR_J}/\mathrm{year}.
\end{equation}
Orbital states record $(a/R_J,e,\lambda,\varpi)$, with mean longitude $\lambda$
and longitude of periapse $\varpi$ in radians. Disk profiles are accompanied by
signed torques, one-sided fluxes, local densities and conservation budgets.
The isolated calibrated run has 498 disk profiles and 10,001 orbital samples;
the buoyancy run has \BuoySnapshots\ profiles and \BuoyOrbitStates\ orbital samples.

The pressure-adjusted $2H_{\rm iso}$ Ganymede calculation ends at
$t=403.1731123385396$ yr and $a=9.568528262004177R_J$, with 640 disk profiles
and 19,881 accepted orbital samples. The corresponding pair calculation ends
at $t=471.8383588721317$ yr, with 1,562 disk profiles and 32,810 orbital samples.
Both were stopped for computational cost through normal saving, without changing
the physical prescriptions or timestep controls to obtain these endpoints.
Boundary-selection and optimizer corrections in the pair calculation are
specified in Appendix~\ref{app:pressure_edges}. The $H_{\rm iso}$ Ganymede
case stopped numerically at 18.726503233580434 yr; the outer-window failure is
also described there.

The early-onset stress test ends at the $3.5R_J$ orbital guard at 935.141024 yr,
with 470 paired disk/budget records and 503,413 orbital states. Continuations
retain the gas density, Cartesian orbit, previous density and timestep,
multistep history, original inventories and cumulative conservation budgets.
The buoyancy calculation additionally retains the current and previous accepted
buoyancy angular momentum. No continuation replenishes the disk or resets its
conservation reference.

All eight histories begin from the undepleted profile in equation~\eqref{eq:disk}.
The historical inviscid coefficient check is used only in
Section~\ref{sec:rafikov}; it does not validate the present modal evolution.
The companion manuscript \citep{MosqueiraBuoyancy2026} treats the full buoyancy
theory and retained-heating calculation, distinct from the prescribed-temperature
comparison here.

\subsection{Full-precision numerical inputs and event identification}

The rounded values elsewhere are for presentation. The calculations retain
their original precision. In cgs units the physical constants and
satellite masses are
\begin{align*}
  G&=6.67430\times10^{-8},& k_B&=1.380649\times10^{-16},&
  m_H&=1.6735575\times10^{-24},\\
  M_J&=1.898\times10^{30},& R_J&=7.1492\times10^9,&
  M_G&=1.4823\times10^{26},\\
  M_C&=1.0776\times10^{26},& \gamma&=1.4.&
\end{align*}
With the stated temperature profile and molecular mass $2.3m_H$, the aspect ratios
are $h_{\rm ad}=0.10308959021636653$ and $h_{\rm iso}=h_{\rm ad}/\sqrt{\gamma}$.
The fixed kernel coefficient is $\eta_\star=0.4002226897760582$ and the
deposition-tail width is $w=0.2641692494$. Appendix~\ref{app:tail_definition}
specifies the reference curve, objective and minimization used for the
fixed-profile fit. These are numerical inputs, not physical precision estimates. The separately retained historical inviscid clearing
coefficient is 5.2495.

The outer-depletion-only control first brackets zero between saved times 125.6 and
127.7 yr; its linear estimate is 127.23 yr at $14.113R_J$. The actual calibrated
torque brackets zero between 134.0 and 136.1 yr, with a linear estimate of 135.89
yr. The 0.1-yr orbital record places the first minimum at 135.9 yr. The pair's
first outer-body $e=h$ crossing is bracketed by 288.2--288.3 yr, with linear
estimate 288.21 yr. These values identify the states used by the rounded
presentation.

Net torques, sums, percentages and diagnostic residuals in the text and tables
are calculated from unrounded data; displayed results are rounded independently.
The pair table expresses
remaining drift as $100\dot a/\dot a(0)$ so that small outward rates remain
visible without displaying percentages marginally above 100.

\subsection{Analysis conventions and figure reconstruction}

For least-squares drift fits, we interpolate each orbital history
to equally weighted 0.1-yr samples, including both endpoints. This avoids
weighting the fit by the number of solver steps. This convention is used for
the five baseline/matched cases; early onset uses the final 100 yr. Buoyancy excursions use separate 500--1,000 yr linear detrending, rms
$[N^{-1}\sum_n(a_n-a_{{\rm fit},n})^2]^{1/2}$ and residual maximum minus
minimum, without frequency filtering or smoothing. The same diagnostics are
also evaluated over 400--1,000 and 900--1,000 yr. Here $N$ is the number of equally spaced
samples, $a_n$ the sampled semimajor axis and $a_{{\rm fit},n}$ its fitted linear
trend.

Baseline/matched local depletion divides by $400000/(R/R_J)\ {\rm g\,cm^{-2}}$;
exterior minima use nodes beyond the orbit, excluding the outer boundary. The
clearing test reconstructs all 511 contributions with identical piecewise-linear
interpolation and launch geometry on saved and initial numerical profiles. Sums
match recorded torques within $2\times10^{-13}$ relatively; signed
deficits/excesses verify equation~\eqref{eq:direct_change}. First reversal means
the first nonnegative total torque after a negative record, including buoyancy
where present. Table~\ref{tab:buoyancy_torque} uses that saved state; interpolated
zeros use its negative/nonnegative bracket, whereas minima use every accepted
orbital state. These are profile controls, not separate evolutions.

We identify the pair's crossing using every saved orbital state and shade the
figures from the first $e>h$ state. The preceding 100-yr fit ends at the last
multiple of 10 yr before crossing. Intervals report each body's maximum $e/h$.
Low-mode sums comprise inner $m=2$ and outer $m=1,2$, normalized by side totals
or same-radius initial controls. Late means integrate flux/torque trapezoidally
with interpolated 500 and 1,000 yr endpoints, divided by 500 yr; low-mode
fractions use mean one-sided flux, not near-zero net torque. Recorded buoyancy
reaction plus ordinary torque is checked against total torque at every disk
state.

Table~\ref{tab:buoyancy_mass} integrates the four saved face-flux terms and moving
boundaries of equation~\eqref{eq:bannulus} trapezoidally, reporting the residual
against annular mass change. Late replenishment is $(\Delta M_\nu+\Delta
M_R)/|\Delta M_B|$; the final reference is $4\pi aH_z\Sigma_{\rm init}(a)$ over
the same final limits. This postprocessing follows the coupled trajectory,
separately from the accepted-step global ledger; it does not evolve transport
terms independently.

Figures~\ref{fig:pressure_ganymede}--\ref{fig:pressure_callistos} use disk
profiles nearest 100 yr, 250 yr and the exact endpoint, without radial
interpolation or smoothing. With $r=R/R_J$ and
$s=\Sigma/(2\times10^4\ {\rm g\,cm^{-2}})$, the plotted pressure is
$\Pi/\Pi_{\rm init}(20R_J)=20s/r$; its fixed normalization preserves pressure
extrema. The pressure-case drift and resonance diagnostics are specified in
Appendix~\ref{app:pressure_resonance}.

For the early-onset stress test, depletion ratios use the initial profile at
identical radial nodes. Local values interpolate the evolved and initial columns
separately; threshold radii interpolate their ratio. An exterior trough requires
a local minimum, not merely the first node beyond the orbit.

As a sampling check, fits using every accepted orbital step were compared with
fits on the uniform 0.1-yr grid. The finite-stress 500--1,000 yr slope differs by
0.60\%, while the original-modal, calibrated and pair results are unchanged at
the quoted precision. Calculations and analysis use Python 3.12.14, NumPy 2.3.5,
SciPy 1.18.1 and Matplotlib 3.11.2.

\section{Density dependence of the adopted deposition prescription}
\label{app:deposition_density}

The reduction tracks extended gaps by sampling separated resonances and depositing
beyond each launch and shock, allowing excitation to move toward distant edges
without a shallow-profile stopping condition.

The launch bound, smooth tail and accounting for deposited and escaping angular
momentum preserve that separation. The tail width is calibrated on the
undepleted reference disk. The following uniform-density and local asymptotic
checks concern damped contributions; $m\leq2$ escape and local buoyancy remain
separate.

\subsection{Retained post-shock tail}
\label{app:tail_definition}

After this onset, deposition is spread using a fixed analytic decay shape:
\begin{align}
  T_\sigma(x)&=\int_0^x(1+\sigma u)^{3/4}
  \left|(1+\sigma u)^{-3/2}-1\right|^{3/2}du,\nonumber\\
  U_\sigma(x)&=\frac{\max[T_\sigma(x)/T_\sigma(x_{\rm sh})-1,0]}{w},
  &Q_\sigma(x)&=1-[1+U_\sigma(x)^2]^{-1/4}.
  \label{eq:tail}
\end{align}
Here $u$ is a dummy dimensionless radial offset, $T_\sigma$ a dimensionless
geometric propagation coordinate and $U_\sigma$ its scaled post-shock increment;
$T_\sigma$ is distinct from temperature $T(R)$. The cumulative deposited
fraction $Q_\sigma$, distinct from the disk stability parameter $Q$, gives
grid-interval deposition by differencing. The surviving fraction approaches
$U_\sigma^{-1/2}$ at large $U_\sigma$, retaining the late-decay asymptote of
\citet{RafikovPropagation2002}. Equation~\eqref{eq:tail} is an analytic surrogate
used in the reference implementation, not an exact published solution for the
whole decay curve. Ginzburg \& Sari instead deposit at the shock without an
extended tail.

The common width was fitted once on a smooth control with Ganymede at $20R_J$,
$\Sigma=2\times10^4\ {\rm g\,cm^{-2}}$ and the stated constant $h$, on 801
uniform nodes from $2$ to $70R_J$. The reference cumulative fraction is
$Q_{\rm ref}=1-\phi(I_{\rm ref})$, where
\begin{equation}
  I_{\rm ref}(R)=\frac{9\mu}{4\,2^{1/4}h^{11/2}}T_\sigma(x),\qquad
  \phi(I)=\left[1+\max(I/0.79-1,0)^2\right]^{-1/4}.
  \label{eq:tail_reference}
\end{equation}
Here $\sigma$ selects the side of $R$. The saved reference evaluates this
coordinate by trapezoidal integration of $dI_{\rm ref}/d|R-a|$ away from the
orbit on each side; at the
two nodes bracketing $a$, it starts from
$I_{\rm ref}=(2/5)|R-a|(dI_{\rm ref}/d|R-a|)$ evaluated at that node. This
specifies the original near-orbit quadrature as well as the continuum formula.

Let $\overline Q_\sigma(R;w)$ be the cumulative fraction obtained by weighting
the individual modal fractions by $A_{m,\sigma}$ and dividing by the full
one-sided launch sum. For the sets $\mathcal J_\sigma$ of radial nodes strictly
inside or outside $a$, with counts $N_\sigma$, the fitted objective was
\begin{equation}
  \mathcal E(w)=\frac12\sum_{\sigma=\pm1}\frac1{N_\sigma}
  \sum_{j\in\mathcal J_\sigma}
  [\overline Q_\sigma(R_j;w)-Q_{\rm ref}(R_j)]^2.
  \label{eq:tail_fit}
\end{equation}
Every node on a given side has equal weight, and the two sides have equal total
weight, over their full represented domains. Both curves use identical launch
sums; no finite-domain renormalization is applied. The fit includes all 511
unsoftened contributions, with no imposed low-mode escape cutoff; modes whose
shock lies outside the domain deposit zero there. It predates the $m\leq2$
escape choice and the $R^{-1}$ disk, and was not repeated for either or for the
3D and buoyancy comparisons. Bounded scalar minimization over
$0.05\leq w\leq5$ with absolute width tolerance $10^{-8}$ gives
$w=0.264$ and $\mathcal E=0.00307$. The inner/outer
rms cumulative differences are 0.0565/0.0544; their maxima are 0.121/0.230.
This is approximate agreement of deposition profiles, not a fit to migration or
a stopping radius.

\subsection{Retained density dependence and uniform scaling}

For one harmonic and side at fixed satellite mass, orbit and temperature, the
positive launched flux is $A=\mathcal C\Sigma_{\rm launch}$, where $\mathcal C$
includes the modal coefficient and any finite-thickness factor. The density ratio
$\Sigma_{\rm sh}/\Sigma_{\rm launch}$ also enters the shock estimate in
equation~\eqref{eq:shock}. In calibrated cases its denominator uses $\mu_{\rm
eff}^2=\mathcal F_{m,\sigma}\mu^2$, as specified in
equation~\eqref{eq:3dshockchoice}. Both the launch density and the receiving
density that selects the shock therefore affect deposition.

Beyond that shock, equation~\eqref{eq:tail} gives the remaining flux magnitude
and, on the outer side, the deposition density
\begin{equation}
  F_{\rm rem}=A(1+U_+^2)^{-1/4},\qquad
  D_{\rm current}=\frac{A\,T_+'(x)}{2awT_+(x_{\rm sh})}
  U_+(1+U_+^2)^{-5/4},\qquad x>x_{\rm sh}.
  \label{eq:tail_density_current}
\end{equation}
Here the prime differentiates the dimensionless geometric coordinate with respect
to $x=|R-a|/a$; the inner contribution has negative gas torque. There is no
deposition before the selected shock. Density enters this expression through $A$
and $x_{\rm sh}$, with no separate dependence on each subsequent receiving cell.

Multiplying the entire sampled density profile, including any prescribed exterior
continuation, by a positive constant $f_\Sigma$ leaves the shock-selection equation
unchanged. Consequently,
\begin{equation}
  \Sigma\mapsto f_\Sigma\Sigma:\qquad
  A\mapsto f_\Sigma A,\quad x_{\rm sh}\mapsto x_{\rm sh},\quad
  Q_\sigma\mapsto Q_\sigma,\quad D\mapsto f_\Sigma D.
  \label{eq:tail_uniform_scaling}
\end{equation}
This is the correct uniform-density homogeneity at fixed orbit and thermal
structure. It does not imply that entire migrating histories, with transport and
fixed boundaries, follow a time rescaling. The approximation concerns spatial
density contrast, rather than a missing overall power of density.

\subsection{Propagation along a nonuniform receiving disk}

In the weakly nonlinear propagation theory of \citet{RafikovPropagation2002},
equations 32--34, the nonlinear coordinate retains the density encountered along
the path. Suppressing fixed normalization factors, write
\begin{equation}
  \frac{dI}{ds}=K(s)\sqrt{\frac{\Sigma_{\rm src}}{\Sigma[R(s)]}},\qquad
  |D_{\rm Rafikov}|=-A\phi'(I)\frac{dI}{ds}.
  \label{eq:tail_density_path}
\end{equation}
The distance $s$ increases along propagation on either side; $K>0$ contains the
sound-speed, shear, geometry and fixed forcing factors, and $\phi$ is the
surviving-flux fraction. The aggregate-wake source normalization is $\Sigma_{\rm
src}=\Sigma(a)$. This form includes both the local density factor and its
cumulative contribution to $I$; it is not an independently evolved wave field in
our secular model.

Consider two profiles with identical density at a mode's launch and
selected shock, but different density farther along its path, with the shock
selection unchanged. Our prescription assigns that contribution the same $A$,
$x_{\rm sh}$ and deposition in both profiles.
Equation~\eqref{eq:tail_density_path} generally does not. A changed interval
affects the accumulated coordinate even after propagation reaches gas whose
density is unchanged. This counterexample isolates the omitted dependence without
altering excitation.

Part of the density scaling is nevertheless recovered by the moving shock. In the
local limit $x\ll1$, $T_\sigma(x)\propto x^{5/2}$. Suppose the receiving-to-launch
density ratio $f=\Sigma_{\rm receiver}/\Sigma_{\rm launch}$ is approximately
constant over the relevant propagation region and the launch bound is inactive. At
fixed $A$, $x_0$, $h$ and $\mu_{\rm eff}$,
\begin{equation}
  x_{\rm sh}\propto f^{1/5},\qquad T_\sigma(x_{\rm sh})\propto f^{1/2},\qquad
  \frac{T_\sigma(x)}{T_\sigma(x_{\rm sh})}\propto f^{-1/2}.
  \label{eq:tail_shock_scaling}
\end{equation}
Far into the adopted tail, where $U_\sigma\gg1$,
\begin{equation}
  F_{\rm rem}\simeq A\left[\frac{wT_\sigma(x_{\rm sh})}{T_\sigma(x)}\right]^{1/2}
  \propto A f^{1/4}x^{-5/4}.
  \label{eq:tail_far_scaling}
\end{equation}
The late weak-shock flux scales as $I^{-1/2}$ \citep{RafikovPropagation2002}; with
$I\propto f^{-1/2}x^{5/2}$ this gives the same $f^{1/4}$ dependence. This is a
limiting algebraic correspondence, not agreement of the full normalization or
deposition shape. Once $x_{\rm sh}=x_0$, further receiving-density reductions
cannot shift the onset, so this correspondence no longer follows.

The distinction also exists on the smooth initial disk. For $c_{\rm ad}\propto R^{-1/2}$
and $\Sigma\propto R^{-p}$, Rafikov's global power-law coordinate (his equation
43), expressed with $q=R/a$, has geometric weight proportional to
\begin{equation}
  q^{3/4+p/2}|q^{-3/2}-1|^{3/2}.
  \label{eq:tail_powerlaw_weight}
\end{equation}
Our tail retains the $p=0$ weight, whereas $p=1$ gives exponent $5/4$. The width
was calibrated on a uniform-density reference. The shock estimate already responds
to the initial gradient, so this difference alone is not an estimate of the total
error on the adopted $R^{-1}$ disk.

\subsection{Conservation, interpretation and a possible diagnostic}

The deposited-plus-escaping identity in equation~\eqref{eq:escape} remains
intact. The uncertainty concerns where that angular momentum is deposited.
Across a structured gap, faster damping in a depleted interval can leave less
flux for subsequent intervals; no universal sign follows for the change in
clearing or migration. At fixed $A$ and $x_{\rm sh}$, a depleted receiving cell
can retain finite prescribed $D$, with specific torque $D/(2\pi R\Sigma)$.
Multiplying that deposition by a local density factor would not restore the
accumulated propagation. Neither this tail nor a weak-shock reduction is a
controlled model of propagation through a true vacuum or arbitrarily abrupt
structure.

The comparable Rafikov clearing scale is a consistency check, not a path-density
test: all eight histories share this tail. Its replacement could change subsequent
disks even though excitation on a fixed profile is unchanged. The isolated
reference Ganymede runs with instantaneous adjustment or $\chi=0.5$ complete
1,000 yr with confinement; $\chi=1$ reaches its
orbital guard with a cavity. Positive density and reported budgets
(Appendix~\ref{sec:conservation}) check those integrations, not spatial convergence
or local deposition accuracy. Exact uniform scaling and the restricted
delayed-shock correspondence hold, but the quantitative influence of full path
dependence remains unestablished.

A future diagnostic on frozen snapshots can compare the density-weighted and
geometric post-shock increments for contribution $j$:
\begin{equation}
  \mathcal C_j(s)=
  \frac{\displaystyle\int_{s_{{\rm sh},j}}^s K_j(s')
  \sqrt{\Sigma(R_{{\rm sh},j})/\Sigma[R_j(s')]}\,ds'}
  {\displaystyle\int_{s_{{\rm sh},j}}^s K_j(s')\,ds'},\qquad s>s_{{\rm sh},j}.
  \label{eq:tail_path_diagnostic}
\end{equation}
It equals one on a constant-density receiving path. Evaluation should focus on
portions carrying appreciable flux and report the deposition share from
contributions with $x_{\rm sh}=x_0$. Here $s_{{\rm sh},j}$ is the shock's
propagation distance, $R_{{\rm sh},j}$ its radius and $R_j(s)$ the radius along
contribution $j$'s path. Initial, reversal and late snapshots would
distinguish evolving path structure without changing accepted histories. These
diagnostics remain unevaluated; no percentage error on a stall radius or gap depth
is assigned.

\section{Conservative Rayleigh adjustment and numerical integration}
\label{app:rayleigh}

\subsection{Marginal stability and compatible fluxes}

The rapid-relaxation assumption in Section~\ref{sec:transport} is implemented
through an auxiliary stress in the retained Keplerian transport equations. The
pressure-balanced rotation supplies the stability diagnostic only.

With $y=\ln(\Sigma/\Sigma_{\rm ref})$, constant
$\Sigma_{\rm ref}=2\times10^4\ {\rm g\,cm^{-2}}$ and $b=h^2/\gamma$, primes
below denote derivatives with respect to $R$. The pressure-balanced angular
frequency $\Omega_{\rm pb}$ and squared radial epicyclic frequency
$\kappa_{\rm pb}^2=R^{-3}d(R^4\Omega_{\rm pb}^2)/dR$ give the diagnostic
\begin{equation}
  \frac{\Omega_{\rm pb}^2}{\Omega_K^2}=1+b(Ry'-1),\qquad
  K[\Sigma]\equiv\frac{\kappa_{\rm pb}^2}{\Omega_K^2}
  =1+b(2Ry'+R^2y''-1).
  \label{eq:stability}
\end{equation}
Thus $K[\Sigma]$ is normalized by $\Omega_K^2$, whereas the finite-transport
indicator $\chi$ in Section~\ref{sec:gradient_sensitivity} uses
$\Omega_{\rm pb}^2$.
The additional stress satisfies the marginal-stability conditions
\begin{equation}
  G_R\geq0,\qquad K[\Sigma]\geq0,\qquad G_RK[\Sigma]=0.
  \label{eq:complementarity}
\end{equation}
Density and stress are solved together at each secular step. Stress vanishes in
strictly stable cells and may persist at marginality under continued forcing.
Iterations enforce instantaneous local adjustment; they do not resolve instability
growth, saturation or a physical relaxation time, or predict a turbulent
viscosity.

For adjacent evolved cells the added mass flux is
\begin{equation}
  F^R_{M,j+1/2}=-\frac{G_{R,j+1}-G_{R,j}}{\ell_{K,j+1}-\ell_{K,j}},\qquad
  \ell_{K,j}F^R_{M,j+1/2}+G_{R,j}
  =\ell_{K,j+1}F^R_{M,j+1/2}+G_{R,j+1}.
  \label{eq:sharedflux}
\end{equation}
The shared face flux cancels interior transfers in the Keplerian inventory.
Boundary fluxes use the actual boundary-to-cell spacing. Added stress is zero in
the two boundary-adjacent evolved cells, which are excluded from the curvature
test; its boundary flux therefore vanishes while ordinary drainage remains active.
Pressure-balanced rotation supplies only the stability test, not an evolved
angular-momentum inventory.

\subsection{Gas and orbital integration}
\label{app:integration}

We solve for $\ln(\Sigma/\Sigma_{\rm ref})$ to keep the gas density positive,
with $\Sigma_{\rm ref}=2\times10^4\ {\rm g\,cm^{-2}}$. The first gas step uses
backward Euler and subsequent steps variable-step BDF2. In the instantaneous
baseline, a semismooth Newton solve enforces the gas residual and the stress
complementarity condition. It uses the previous accepted stress only as a
starting guess. Both three-point and five-point curvature estimates must satisfy
the chosen Rayleigh criterion on nodes $2$ through $N_R-3$, where $N_R=801$
is the gas-node count (zero-based indexing). On these nodes,
$\Omega_{\rm pb}^2$ must also remain positive. The maximum iteration count
is 70, the scaled residual tolerance $2\times10^{-11}$ and the allowed
dimensionless stability residual $2\times10^{-7}$. No density floor, clipped
stress or fallback density reset is used. The finite-transport comparison uses
the same log-density variable, gas time integration and residual tolerance, with
the stress given directly by equation~\eqref{eq:gradient_alpha} and its
viscosity relation instead of complementarity. Positive $\Omega_{\rm pb}^2$
remains required.

The gas step satisfies
\begin{equation}
  \Delta t\leq\min\left[0.1\ {\rm yr},\frac{0.1\Delta R}
  {\max_i|2\Gamma_i/(M_i a_i\Omega_i)|}\right],
  \qquad \Delta R=0.085R_J.
  \label{eq:timestep}
\end{equation}
Source locations are predicted at the half step. The gas density and Rayleigh
stress are solved at those locations; Cartesian orbits then advance with the
average old and new gas torque. The eccentricity-damping time is evaluated using
the mean old/new density at the old orbit. The source locations are rebuilt at the
accepted new semimajor axes. An unsuccessful nonlinear step halves $\Delta t$
without accepting the trial state. A shock at a grid knot retains its original
bracketing interval when differentiating the interpolated density.

Let $\boldsymbol r_i$ and $\boldsymbol v_i$ be planetocentric position and
velocity, $r_i=|\boldsymbol r_i|$ and $\widehat{\boldsymbol z}$ the disk-normal
unit vector. The orbital calculation integrates direct mutual gravity, the indirect
acceleration of the planet-centered frame, central gravity with $G(M_J+M_i)$,
tangential acceleration $\Gamma_i\widehat{\boldsymbol z}\times\boldsymbol r_i/(M_i
r_i^2)$, and radial eccentricity damping $-2(\boldsymbol r_i\cdot\boldsymbol
v_i)\boldsymbol r_i/(t_{e,i}r_i^2)$. The latter contributes no direct orbital
angular momentum. The adopted reference damping time uses the coefficient of
\citet{TanakaWard2004}:
\begin{equation}
  t_{e,i}=\frac{1}{0.780}\frac{M_J}{M_i}
  \frac{M_J}{\Sigma(a_i)a_i^2}h^4\Omega_K^{-1}(a_i).
  \label{eq:edamp}
\end{equation}
Here the retained code uses $h=h_{\rm ad}$. Tanaka \& Ward's isothermal
calculation uses $h_{\rm iso}$; at fixed temperature this adopted reference time
is larger by $\gamma^2$. We state that convention as part of the circular-wave
closure, rather than interpreting it as a new adiabatic eccentricity-damping
calibration.
There is no imposed resonant lock or stopping radius. A fourth-order Runge--Kutta
integrator uses substeps limited by $P_{\min}/400$ and
$0.02[d_{ij}^3/G(M_i+M_j)]^{1/2}$, where $P_{\min}$ is the shortest
instantaneous osculating orbital period and
$d_{ij}=|\boldsymbol r_i-\boldsymbol r_j|$ is the satellite separation.
Mutual forces are unsoftened.

Encounter, invalid-element and substep-limit guards stop an orbital advance
if any satellite separation is $d_{ij}<0.0673R_J$, a semimajor axis is
nonpositive or nonfinite, or more than $10^6$ substeps are required.
Other stops are $a\notin(3.5,65)R_J$,
$\min(\Sigma/\Sigma_{\rm ref})<10^{-10}$, or $\Delta t<10^{-7}$ yr. The density
threshold is a guard, not a floor.

\subsection{Conservative pressure-bump adjustment}
\label{app:pressure_edges}

Section~\ref{sec:pressure_edges} adds a zero-time redistribution after each
accepted secular-step candidate. Its pressure trigger is independent of a
Rayleigh violation and can act on a Rayleigh-stable edge.
A pressure maximum, including a flat-topped
maximum, is identified by a rise followed by a fall in $\ln\Pi$. Adjacent
changes no larger than $2\times10^{-9}$ are treated as numerical ties. For a
maximum at $R_p$, the window is $R_p\pm w_P H_{\rm iso}(R_p)$, with dimensionless
half-width $w_P=1$ or 2, distinct from the deposition-tail width $w$.
Overlapping windows are merged, so the actual adjusted
region can exceed one nominal window. A maximum whose rising flank connects to
the first evolved cell identifies the drain feature; its entire connected group
of overlapping windows is excluded. In the isolated histories this connection
requires no fall larger than the detector tolerance along that flank. In the
pair continuation, tiny reversals in the boundary profile required a net-rise
criterion: only the innermost overlapping-window group is excluded when its
highest $\ln\Pi$ exceeds that at grid node 2 by more than $2\times10^{-9}$.
Node numbering is zero-based. Detached outer groups remain eligible. This
avoids reclassifying the drain feature as a gap edge; the original draining
boundary and Rayleigh transport remain active. Windows that do not fit within
the evolved domain are skipped and recorded.

Let $W$ be the set of cells in the selected window,
$M_j=2\pi R_j\Delta R\Sigma_j$ the pre-adjustment cell mass, $M_W$
the window mass, and $z_j=\ln(\Sigma_j^{\rm new}/\Sigma_j)$.
We minimize
\begin{equation}
  \frac12\sum_{j\in W}\frac{M_j}{M_W}z_j^2,
  \qquad
  \sum_{j\in W}M_j(e^{z_j}-1)=0,\qquad
  \sum_{j\in W}M_j\ell_{K,j}(e^{z_j}-1)=0,
  \label{eq:pressure_projection}
\end{equation}
subject to monotonic pressure within the window, positive density and the
retained three- and five-point Rayleigh constraints, including adjoining cells.
For the prescribed $T\propto R^{-1}$, constant pressure requires
$\Sigma\propto R$. Its mean specific angular momentum on this uniform grid is
\begin{equation}
  \overline\ell_{\Pi={\rm const}}
  =\frac{\sum_{j\in W}R_j^2\ell_{K,j}}{\sum_{j\in W}R_j^2}.
  \label{eq:constant_pressure_moment}
\end{equation}
Pressure is required to be nonincreasing when the conserved window mean
$\sum_{j\in W}M_j\ell_{K,j}/M_W$ does not exceed this reference, and
nondecreasing otherwise. This permits sloping portions and plateaus; a uniform
pressure throughout a fixed window generally cannot preserve both inventories.

For $\Delta M_j=M_j(e^{z_j}-1)$, define the integrated outward mass transfer
$\mathcal I_{M,j+1/2}$ and integrated auxiliary stress $\mathcal G_j$ by
\begin{equation}
  \mathcal I_{M,j+1/2}=-\sum_{k\leq j}\Delta M_k,
  \qquad
  \mathcal G_{j+1}-\mathcal G_j
  =-\mathcal I_{M,j+1/2}(\ell_{K,j+1}-\ell_{K,j}).
  \label{eq:pressure_impulse}
\end{equation}
The sum begins at the window's first cell. The first stress is zero; the two
inventory constraints close the boundary mass transfer and terminal stress.
We require $\mathcal G_j\geq0$ and verify the compatible flux reconstruction.
Gas outside the window is unchanged. The integrated stress
$\mathcal G_j=\int G_{P,j}\,dt$ has units of angular momentum, whereas the
auxiliary torque $G_{P,j}$ has the same units as $G_R$. This integral denotes
the zero-time transfer, not a predicted turbulent viscosity or relaxation time.

Nonnegative integrated stress also gives the dissipative sign for the represented
circular orbital energy. Define $E_K=\sum_{j\in W}M_j\varepsilon_j$, with
$\varepsilon_j=-GM_J/(2R_j)$, and the secant frequency
\begin{equation}
  \overline\Omega_{j+1/2}
  =\frac{\varepsilon_{j+1}-\varepsilon_j}{\ell_{K,j+1}-\ell_{K,j}}.
  \label{eq:pressure_energy_secant}
\end{equation}
With zero boundary transfers and endpoint stresses, summation by parts gives
\begin{equation}
  \Delta E_K=\sum_{j\in W^{\circ}}\mathcal G_j
  (\overline\Omega_{j+1/2}-\overline\Omega_{j-1/2})\leq0,
  \label{eq:pressure_energy_sign}
\end{equation}
where $W^{\circ}$ excludes the two endpoint cells. The secants decrease outward
because $d\varepsilon/d\ell_K=\Omega_K$ decreases outward. This property holds
for the exact constraints; the numerical acceptance tolerances are given below.
It concerns circular orbital energy, not the full gas energy under prescribed
temperature.

The constrained solve uses sequential least-squares programming (SLSQP), with
objective tolerance $2\times10^{-13}$,
150 iterations and search bounds $-30\leq z_j\leq30$. Acceptance independently
checks relative mass and centered-angular-momentum residuals below
$2\times10^{-11}$, log-pressure ordering to $2\times10^{-9}$, both Rayleigh
diagnostics to $2\times10^{-7}$, and positive pressure-supported rotation.
The optimizer's Rayleigh target is $K\geq-10^{-9}$. The allowed negative
integrated-stress residual is $2\times10^{-11}M_W\Delta\ell_K$, where
$\Delta\ell_K$ spans the window; the density change reconstructed from its
fluxes must agree to $2\times10^{-9}$ times the largest input column in the
window. There is no density floor or nonconservative fallback. A failed
adjustment rejects the entire trial step and halves its timestep under the
existing guard.

The ordinary gas and orbital step uses its pre-adjustment endpoint. The pressure
impulse then changes the gas profile without adding orbital torque. After a
nonzero impulse the BDF history is cleared and the next step uses backward
Euler; this additional operator splitting is first order. Original conservation
inventories and accumulated boundary exchanges are retained across checkpoint
continuations. Satellite forcing continually regenerates the pressure bumps,
so the adjustment acts repeatedly even when saved post-adjustment profiles
contain no resolved bump. Resolving that repeated rebuilding and redistribution
requires small accepted timesteps and slows the computation. These timesteps
do not represent a resolved physical instability growth or relaxation time.
Both $2H_{\rm iso}$ histories were stopped for computational cost after normal saving.
These endpoints do not represent breakdown of the redistribution operator.

Pressure maxima are checked independently after redistribution, including
features shifted to window boundaries. For each retained maximum, its prominence
is its pressure above the higher of the two surrounding trough levels obtained
before encountering a higher peak or the domain boundary, divided by the peak pressure. Numerical
plateaus use their strict grid maxima for this calculation. Depletion intervals
are contiguous threshold crossings around the exterior minimum of
$\Sigma(R,t)/\Sigma(R,0)$, with linearly interpolated endpoints. These are
saved-profile diagnostics, not a guarantee that no transient feature forms
between outputs.

The pair uses the following documented sequence of boundary-selection rules.
Initially the drain exclusion required a rising flank from node 1. At
252.196 yr its starting point moved to node 2, excluding the boundary-adjacent
face; at 282.941 yr it changed to the net-rise rule for the innermost window
group described above. These changes affect selection, separately from the
subsequent optimizer rescaling. Comparing the earlier selectors on all 156
saved pre-adjustment profiles through the latter correction gives the same
windows as the active rule and the original saved window records. On the two
rejected drain-trial fixtures, the revised rules remove only the drain window,
retaining the gap windows. This checks the saved states, not every intervening
accepted step. Applying the final detector retrospectively does not re-evolve
the earlier history. A subsequent gap-window adjustment at the latter epoch
required numerical scaling to resolve nearly tied pressure values. For this continuation the optimizer uses $u_j=z_j/s$, where
$s$ is the largest violation of the selected log-pressure ordering, clipped to
$10^{-8}\leq s\leq1$. Its objective is divided by $s^2$; its inventory and
stress constraints are evaluated with $\operatorname{expm1}(su_j)$ and scaled
by $s$. The ordering target allows $1.8\times10^{-9}$ per face, inside the
unchanged $2\times10^{-9}$ detector and final acceptance tolerance. Rayleigh,
inventory, positive-density and flux acceptance checks are unchanged. The
isolated histories use the original unscaled optimization. No rejected trial
density is accepted into the history.

The stored windows and integrated stresses specify the region actually
redistributed at each saved step; Section~\ref{sec:pressure_criterion} gives the
final merged-window example. The selector comparison and a manufactured check
of equation~\eqref{eq:pressure_energy_sign} use fixed inputs without evolving
the disk.

The $H_{\rm iso}$ case stopped numerically at 18.7 yr, with Ganymede at
$19.1R_J$, when adjustment in the outer window $21.6$--$25.6R_J$ could not be
resolved. It does not supply a late-time width comparison. The completed
$2H_{\rm iso}$ Ganymede case has maximum saved mass and added-transport
angular-momentum residuals of $1.71\times10^{-11}$ and $9.66\times10^{-12}$
of the original disk inventories. Its full signed residual is $-0.102\%$ of
the absolute orbital angular-momentum change. The corresponding pair values
are $4.32\times10^{-12}$, $3.16\times10^{-12}$ and $+0.0277\%$.
The pair's pressure diagnostic applies its final boundary-selection rule to
all saved profiles; the original per-step classifications are also retained.
Numerical specifications and analysis conventions are in
Appendix~\ref{app:reproduction}.

\subsection{Resonance and drift diagnostics for the pressure-adjusted pair}
\label{app:pressure_resonance}

The three second-order 7:5 resonant angles are
\begin{align}
  \theta_1&=7\lambda_{\rm out}-5\lambda_{\rm in}-2\varpi_{\rm in},\nonumber\\*
  \theta_2&=7\lambda_{\rm out}-5\lambda_{\rm in}
                  -\varpi_{\rm in}-\varpi_{\rm out},\nonumber\\*
  \theta_3&=7\lambda_{\rm out}-5\lambda_{\rm in}-2\varpi_{\rm out}.
  \label{eq:pair_pressure_angles}
\end{align}
They obey $\theta_1=\theta_2+\Delta\varpi$ and
$\theta_3=\theta_2-\Delta\varpi$, modulo $2\pi$. Thus bounded libration
of $\theta_2$ and $\Delta\varpi$ also implies bounded libration of
$\theta_1$ and $\theta_3$.
All three angles are reconstructed from the saved mean longitudes and
periapse longitudes.
We form each combination before wrapping or unwrapping it. All accepted-step
orbital samples are retained; the largest sampling interval over the checked
200-yr-to-endpoint window is 0.1 yr. Each unwrapped angle spans less than
$61^\circ$ over that interval. Over 400 yr to the endpoint, their half
peak-to-peak excursions are $17.5^\circ$, $17.4^\circ$ and $17.2^\circ$,
about time-weighted circular centers $183^\circ$, $1.76^\circ$ and
$180^\circ$. Trapezoidal time weights avoid overweighting the dense late
sampling. The apsidal difference centers on $181^\circ$, with a
$0.283^\circ$ half peak-to-peak excursion. These bounded angles establish the
reported resonance over the checked interval; its initial capture time is not
assigned.

Figure~\ref{fig:pressure_resonance} plots $\theta_2$ and
$\Delta\varpi$ over the full accepted orbital history and a 400--410 yr
detail. The resonant angle is wrapped to $[-180^\circ,180^\circ]$ and the
apsidal separation to $[0^\circ,360^\circ)$, with no subtracted offsets.
Lines are broken at wrapping jumps; no phase or time smoothing is applied. Angle samples with either eccentricity no larger than
$10^{-8}$ are omitted because periapse is undefined for a circular orbit.
The eccentricity panel retains every sample. Shading identifies the checked
200-yr-to-endpoint interval, not a fitted capture time.

The common 400--470 yr drift comparison uses the uniform 0.1-yr interpolation
and least-squares convention of Section~\ref{sec:numerics}. The final 10- and
100-yr mean rates instead use the actual endpoint displacement divided by the
interval duration, interpolating the starting orbit from the accepted samples. They include
mutual orbital exchange and are not the individual instantaneous disk-torque
migration rates. The intermoon mean depletion is the current annular gas mass
divided by the initial mass over the same radial nodes between the two
semimajor axes. The quoted trough fraction compares the two columns at its
single radius.

\subsection{Conservation across the six reference histories}
\label{sec:conservation}

The six histories here exclude the two pressure-adjustment cases in
Table~\ref{tab:case_map}. The budgets of those two additional histories are
reported in Appendix~\ref{app:pressure_edges}.
Table~\ref{tab:budgets} distinguishes the conservative Rayleigh transport from the
residual of the complete coupled calculation. The added operator's accumulated
residual is of order $10^{-11}$ of the initial disk angular momentum in the five
baseline and matched histories, and $4.11\times10^{-8}$ in the early-onset stress
test with more than $5\times10^5$ accepted gas steps. The full residual
additionally includes ordinary spatial discretization, boundary quadrature and
disk--orbit time coupling; it is reported relative to the absolute orbital
angular-momentum change. These evolutions do not use the earlier nonconservative
density reset.

In the table, $M_0$ and $J_{d,0}$ are the initial evolved gas mass and
Keplerian disk angular momentum. The mass residual $\delta M$ is the gas-mass
change plus cumulative net outward boundary mass exchange. The signed
$\Delta J_s$ is the change in barycentric orbital angular momentum, including
the planetary reflex correction. The residual $\delta J_R$ is the disk
angular-momentum change minus the accepted-step ledger of represented gas
transport, including the ordinary operator, buoyancy when present and the
added-stress boundary contribution; it is not the angular momentum physically
carried by the added stress. The full residual $\delta J_{\rm full}$ sums the
disk and orbital changes with cumulative net outward gas angular-momentum
flux (advection and boundary stress) and signed wave escape. Cumulative
exchanges use the accepted backward-Euler/BDF2 weights.

\begin{table}[!htbp]
  \centering
  \small
  \setlength{\tabcolsep}{3pt}
  \caption{Conservation summary for the six reference histories, with residuals and
  normalizations defined above. Accepted and rejected counts refer to gas steps,
  excluding the final orbital guard trial. Maxima use saved budget records;
  the full residual is the signed endpoint percentage. Boundary exchanges and
  signed modal escape are included. The two angular-momentum residuals have
  different normalizations.}
  \label{tab:budgets}
  \begin{tabular}{lrrrrr}\toprule
Case & Accepted & Rejected & $\max|\delta M|/M_0$ & $\max|\delta J_R|/J_{d,0}$ & $\delta J_{\rm full}/|\Delta J_s|$\\\midrule
Ganymede: original modal & 10,000 & 0 & $3.47\times10^{-11}$ & $2.39\times10^{-11}$ & $-0.250\%$ \\
Ganymede: calibrated 3D & 10,000 & 0 & $5.74\times10^{-11}$ & $3.94\times10^{-11}$ & $-0.145\%$ \\
Ganymede: finite transport & 21,647 & 3782 & $1.39\times10^{-11}$ & $9.71\times10^{-12}$ & $-0.0957\%$ \\
Ganymede: early onset & 503,412 & 99313 & $7.43\times10^{-8}$ & $4.11\times10^{-8}$ & $+0.0283\%$ \\
Ganymede: with buoyancy & 10,003 & 1 & $1.50\times10^{-11}$ & $1.08\times10^{-11}$ & $-0.0749\%$ \\
Two Callistos & 5,474 & 0 & $2.27\times10^{-11}$ & $1.68\times10^{-11}$ & $-0.0602\%$ \\
\bottomrule\end{tabular}

\end{table}

\FloatBarrier

\end{document}